\documentclass[aps,pra,twocolumn,showpacs,superscriptaddress,floatfix,10pt]{revtex4}
\usepackage[pdftex]{graphicx}% Include figure files
\usepackage{dcolumn}% Align table columns on decimal point
\usepackage{bm}% bold math
\usepackage[usenames, dvipsnames]{color}
\usepackage{comment}

\usepackage[T1]{fontenc}

\usepackage{amsfonts}
\usepackage{amssymb}
\usepackage{amsmath}
\usepackage{amsthm}
\usepackage{mathrsfs}

\usepackage{subfigure}
\usepackage{rotate}
\usepackage{dsfont}
\usepackage{enumitem}

\usepackage{multirow}
\usepackage{array}
\newcolumntype{L}[1]{>{\raggedright\let\newline\\\arraybackslash\hspace{0pt}}m{#1}}
\newcolumntype{C}[1]{>{\centering\let\newline\\\arraybackslash\hspace{0pt}}m{#1}}
\newcolumntype{R}[1]{>{\raggedleft\let\newline\\\arraybackslash\hspace{0pt}}m{#1}}
\usepackage{tabularx}

\newif\ifnotes
\notestrue

\let\oldmarginpar\marginpar
\renewcommand\marginpar[1]{\-\oldmarginpar[\raggedleft\tiny #1]%
{\raggedright\tiny #1}}

\newcommand{\be}{\begin{equation} }
\newcommand{\ee}{\end{equation} }
\newcommand{\ba}{\begin{eqnarray} }
\newcommand{\ea}{\end{eqnarray} }

\newcommand{\bit}{\begin{itemize}}
\newcommand{\eit}{\end{itemize}}

\newcommand{\J}{J}
\newcommand{\I}{\mathrm{i}}
\newcommand{\AJ}{A_{\J}}
\newcommand{\BJ}{\bar{\J}}
\newcommand{\h}{h}

\newcommand{\Ah}{A_{\h}}

\newcommand{\Bh}{\bar{\h}}
\newcommand{\Q}{Q}

\renewcommand{\Re}{\mathrm{Re}\,}
\renewcommand{\Im}{\mathrm{Im}\,}

\newcommand{\e}{\mathrm{e}}

\newcommand{\x}{\sigma^x}
\newcommand{\y}{\sigma^y}
\newcommand{\z}{\sigma^z}

\renewcommand{\d}{\mathrm{d}}

\usepackage{wasysym}

\def\bra#1{\langle#1|}
\def\ket#1{|#1\rangle}
\def\braket#1#2{\langle#1|#2\rangle}
\def\cexp#1{\left[#1\right]}
\def\qexp#1#2{\bra{#2}#1\ket{#2}}
\def\gsexp#1{\langle #1 \rangle}

\def\dev#1#2{\frac{d#1}{d#2}}
\def\pdev#1#2{\frac{\partial#1}{\partial#2}}
\def\tr#1{\mathrm{tr}\left(#1\right)}
\def\LL{\zeta}

\begin{document}

\title{Critical behavior of the quasi-periodic quantum Ising chain}

\author{P. J. D. Crowley}
\email{philip.jd.crowley@gmail.com}
\affiliation{Department of Physics, Boston University, Boston, MA 02215, USA}

\author{C. R. Laumann}
\affiliation{Department of Physics, Boston University, Boston, MA 02215, USA}

\author{A. Chandran}
\affiliation{Department of Physics, Boston University, Boston, MA 02215, USA}

\date{\today}

\begin{abstract}

The interplay of correlated spatial modulation and symmetry breaking leads to quantum critical phenomena intermediate between those of the clean and randomly disordered cases.
By performing a detailed analytic and numerical case study of the quasi-periodically (QP) modulated transverse field Ising chain, we provide evidence for the conjectures of Ref.~\cite{crowley2018quasi} regarding the QP-Ising universality class. 
In the generic case, we confirm that the logarithmic wandering coefficient $w$ governs both the macroscopic critical exponents and the energy-dependent localisation length of the critical excitations.
However, for special values of the phase difference $\Delta$ between the exchange and transverse field couplings, the QP-Ising transition has different properties.
For $\Delta=0$, a generalised Aubry-Andr\'e duality prevents the finite energy excitations from localising despite the presence of logarithmic wandering.
For $\Delta$ such that the fields and couplings are related by a lattice shift, the wandering coefficient $w$ vanishes.
Nonetheless, the presence of small couplings leads to non-trivial exponents and localised excitations.
Our results add to the rich menagerie of quantum Ising transitions in the presence of spatial modulation.

\end{abstract}

\maketitle

\section{Introduction}
\label{sec:intro}

In the vicinity of a quantum Ising phase transition in a spatially homogeneous (clean) system, the magnetisation (the order parameter) fluctuates on the respective macroscopic length and time scales,
\begin{align}
    \xi\sim \delta^{-\nu}, \quad \xi_t \sim \xi^z,
    \label{eq:scalingxi}
\end{align}
where $\nu$ and $z$ are the correlation length and dynamic exponent respectively, and $\delta$ is the control parameter which measures the deviation from the transition~\cite{Goldenfeld:1992aa}. 
These fluctuations of the order parameter are mediated by long wavelength, low energy excitation modes. 
In the clean transverse field Ising model (TFIM) the transition is in the celebrated Onsager universality class with $\nu=z=1$~\cite{onsager1944crystal,suzuki2012quantum}.

\begin{table}
  \renewcommand{\arraystretch}{1.8}
  \begin{ruledtabular}
\begin{tabular}{ll || l | l  l  l | l }
& Case& $w$ & $\nu$ & $z$ & $\gamma$ & $z_{\mathrm{L}}$ \\ 
\hline
\hline
0.& Ising (Weak modulation)    & 0 & $1$     & $1$    & $7/4$      & -- \\
1.& QP-Ising (Generic $\Delta$)    & $1.2$ & $1^{+}$ & $1.9$ & $2.6^{+}$ & $1.9$ \\
2.& Zero-wandering ($\Delta \in 
Q(\mathbb{N}+\tfrac{1}{2})$) & 0 & $1$    & $3/2$  & $2.2$     & $3/2$ \\
3.& Aubry-Andr\'e ($\Delta = 0$)        & $1.5$ & $1^{+}$ & $2.0$ & $2.7^{+}$ & -- \\
\end{tabular}
\end{ruledtabular}
\caption{
\emph{Summary of critical exponents for smooth quasi-periodic Ising transitions}:
The logarithmic wandering coefficient $w$, correlation length exponent $\nu$, dynamical exponent $z$, susceptibility exponent $\gamma$ and localisation length exponent $z_{\mathrm{L}}$ for the model in Eq.~\eqref{Eq:ModelHam} in various regimes.
All data presented for $Q = (1+\sqrt{5})/2$ the golden mean. The exponents for cases 2 \& 3 are obtained here for the first time.
(Case 0) Weak QP modulation is irrelevant to the clean Ising transition. 
(Case 1) Strong QP modulation is generically relevant due to the logarithmic wandering $w > 0$; this enhances $\nu$ logarithmically (indicated by superscript $^+$), modifies $z$ and $\gamma$, and induces localisation of the finite energy modes.
(Case 2) For special relative phases $\Delta=Q(\mathbb{N}+\tfrac{1}{2})$, $w$ vanishes but the weak couplings nonetheless induce localisation. This case violates the conjecture that $w$ controls the macroscopic critical exponents.
(Case 3) For $\Delta = 0$, a generalised Aubry-Andr{\'e}-type duality prevents localisation of the finite energy modes. Nonetheless, the wandering modifies equilibrium exponents.
}
\label{tab:Isingexponents}
\end{table}

Spatial modulation of the couplings can change the universality class of a quantum phase transition. One feature of this is that locally different regions of the system may be closer to, further from, or even on different sides of, the critical point $\delta = 0$. This is quantified by $\delta_i$, the local deviation from the transition point at the spatial position $i$. 
If the fluctuations of the spatially averaged $\delta$ in a region of size $l$ grow sufficiently quickly with $l$, then the clean transition is perturbatively unstable by the Harris-Luck criterion~\cite{Harris:1974aa,Luck:1993ad,Luck:1993fu}.
Accordingly, random modulation destabilises the clean Ising transition and ultimately the system flows to an infinite-randomness critical point~\cite{McCoy:1968aa,mccoy1969theory,Shankar:1987aa,fisher1992random,fisher1995critical,fisher1999phase,motrunich2000infinite}.
Both quasi-periodic and hyper-uniform modulation allow the fluctuations of $\delta$ to be tuned, and can send the system to new fixed points~\cite{tracy1988universality,kolavr1989attractors,benza1990phase,lin1992phase,Luck:1993ad,turban1994surface,grimm1996aperiodic,hermisson1997aperiodic,igloi1997exact,igloi1998random,hermisson1998surface,crowley2018quasi,crowley2018quantum}.
 
For sufficiently strong smooth quasi-periodic (QP) modulation of the couplings in the TFIM, Ref.~\cite{crowley2018quasi} showed that the fluctuations of $S_{l}(j) = \sum_{i=j}^{j+l-1} \delta_i$, the \emph{wandering}, grow logarithmically with region size $l$:
\begin{align}
    \sigma^2(S_l) \sim w \log(l).
    \label{eq:LogWanderIntro}
\end{align}
The logarithmic growth violates the Harris-Luck criterion but not strongly enough to drive the system to infinite randomness.
Ref.~\cite{crowley2018quasi} argued that the resulting QP-Ising transitions belonged to a new line of intermediate fixed points parameterised by $w$.
At the QP-Ising transitions, macroscopic observables obey power-law scaling (as in the clean Ising transition but with distinct scaling data), while the finite energy excitations are localised (as in the disordered Ising transition).
Table~\ref{tab:ModIsing} summarises the critical behaviours of the clean, QP modulated and disordered Ising transitions which have been studied in the literature.
Case 1 in Table~\ref{tab:Isingexponents} provides representative values of the critical exponents for the QP-Ising transition with a specific $w$.

\begin{figure}
\begin{center}
\includegraphics[width=\columnwidth]{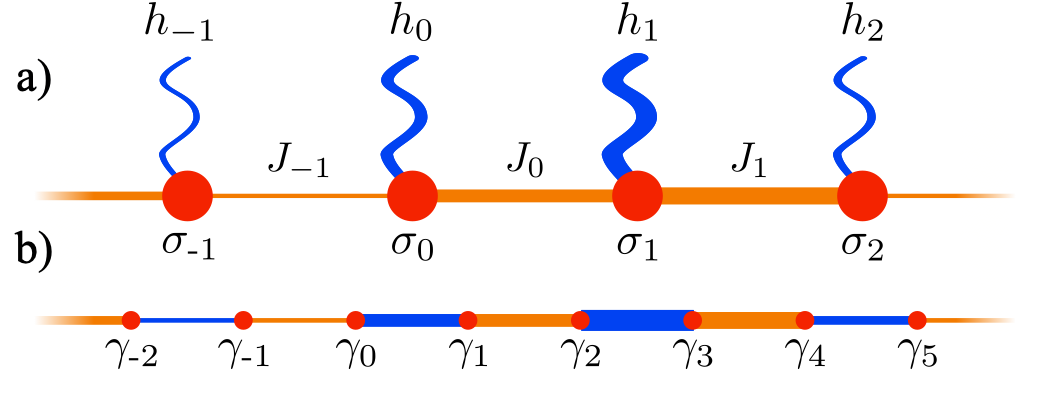}
\caption{
\emph{Quasi-periodically modulated transverse field Ising model}: a) In the spin chain the exchange couplings $\J_j$ and the fields $h_j$ are QP modulated~\eqref{eq:SineMod}. b) The Jordan Wigner transformation maps the spin chain to a chain of non-interacting Majorana fermions. 
}
\label{Fig:IsingChain}
\end{center}
\end{figure}

In this article, we extend the study of Ref.~\cite{crowley2018quasi} and provide evidence in support of two conjectures: 
\begin{enumerate}[label=(\Alph*)]
    \item The logarithmic wandering coefficient $w$ captures the microscopic detail necessary to determine the macroscopic critical exponents of the QP-Ising transition. That is, $w$ parameterises a line of critical fixed points. 
    
    \item The finite energy excitations are localised with a localisation length $\zeta (\epsilon)$ which diverges as $\epsilon \to 0$ with the same dynamical exponent $z_{\mathrm{L}}$ as that governing the equilibrium correlations. Thus, $z_{\mathrm{L}} = z$ up to logarithmic corrections. 
\end{enumerate}
While the two conjectures are generically true at the QP-Ising transition, fine tuning can violate either of them. 
The Zero-Wandering case and Aubry-Andr\'{e} case (Cases 2 and 3 in Table~\ref{tab:Isingexponents}) provide examples of finely tuned models that violate the first and second conjecture respectively. 

A challenge in the study of QP models is to separate physically robust observables from the mathematically intriguing tower of multi-fractal turtles on which they ride.
Our approach is to focus on the macroscopic exponents which govern spatially averaged response and neglect the highly structured scale-dependent fluctuations about the mean trends in any given correlator.
Thus, we supplement a calculation of $\nu$ with measurements of the dynamical exponent $z$ and susceptibility exponent $\gamma$. $z$ sets the low temperature behaviour of the specific heat $c \sim T^{1/z}$ and is extracted from the global density of states $\rho \sim \epsilon^{1/z-1}$, while $\gamma$ controls the divergence of the susceptibility to a longitudinal field $\chi \sim \delta^{-\gamma}$, and is extracted from spatially averaged two-point correlation functions using scaling relations. 

The paper is structured as follows.
We first review the QP-TFIM (Sec.~\ref{sec:prelim}) and its equilibrium properties (Sec.~\ref{sec:phase_diag}). We then calculate the logarithmic wandering $w$ for different values of $\Delta$ and show that strong-smooth modulation violates the Harris-Luck criterion (Sec.~\ref{sec:QPscaling}).
In Sec.~\ref{sec:eqexponents}, we compute $\nu, z$ and $\gamma$ for the QP-Ising transition and provide evidence in support of conjecture A in smooth and square-wave modulated TFIMs.
We also compute the critical exponents for the Zero-wandering and Aubry-Andr\'e transitions, and show that conjecture A is violated for the zero-wandering transition.
In Sec.~\ref{sec:excitations}, we turn to the localisation properties of the Fermionic excitations.
We show that the excitations are localised with $z_{\textrm{L}}=z$ at the zero-wandering and the QP-Ising transitions, but are critically delocalised at the Aubry-Andr\'e transition. 
The Aubry-Andr\'e transition therefore violates conjecture B.
We end in Sec.~\ref{sec:dynamics} with striking dynamical consequences of the localisation for wave-packet spreading.

 %%%%%%%%%%%%%%%%%%%%%%%%%%%%%%

\begin{table*}
  \renewcommand{\arraystretch}{1.8}
\begin{ruledtabular}
\begin{tabular}{ C{65pt} || C{215pt} | C{215pt} }
 & \multicolumn{2}{c}{Spatial structure of low energy excitations} \\

 & Delocalised & Localised  \\ 

\hline
\hline

%\multirow{ 2}{*}{\parbox[t]{2mm}{\multirow{3}{*}{\rotatebox[origin=c]{90}{Universality}}}}&
Ising 
& 
Clean or periodic modulation~\cite{lieb1961two,onsager1944crystal} \newline Weak-continuous-QP modulation~\cite{Luck:1993ad} \newline
Fine-tuned~discontinuous-QP~\cite{doria1988quasiperiodicity,igloi1988quantum,ceccatto1989quasiperiodic,kolavr1989attractors,benza1989quantum,benza1990phase,Luck:1993ad,grimm1996aperiodic,hermisson1997aperiodic,igloi1997exact,igloi1998random,hermisson1998surface}
& 
Strongly hyper-uniform random disorder~\cite{crowley2018quantum} 
\\

\hline

QP-Ising &   \textbf{Aubry-Andr\'{e} strong-continuous-QP modulation ($\Delta = 0$) } \newline Generic-discontinuous-QP modulation & \textbf{Generic strong-continuous-QP modulation ($\Delta \neq 0$),}~(See also~\cite{Chandran:2017ab,crowley2018quasi}) \\

\hline

Infinite \newline randomness & --- & Independent random disorder~\cite{fisher1992random,fisher1995critical,fisher1999phase} \newline Weakly hyper-uniform random disorder~\cite{crowley2018quantum} \\
\end{tabular}
\end{ruledtabular}
\caption{
\emph{Symmetry breaking fixed points of modulated Ising Chains}:
Different modulation organised by universality (rows), and the localisation of low energy excitations (columns). QP modulation is Ising relevant if it is either strong or discontinuous ($\J(\theta)$,$\h(\theta)$ have zeros or jump discontinuities respectively).
In this manuscript we focus on the role of $\Delta$ for strong and smooth QP modulation (bold).
}
\label{tab:ModIsing}
\end{table*}

\section{Preliminaries}
\label{sec:prelim}
\subsection{Model}

The Hamiltonian of the one-dimensional QP-TFIM~\cite{Chandran:2017ab,crowley2018quasi} is
\begin{align}
\label{Eq:ModelHam}
H &= -\frac{1}{2} \left( \sum_{j=1}^{L-1}  \J_j \sigma_j^x \sigma_{j+1}^x + \sum_{j=1}^{L}  \h_j \sigma_j^z\right) -  B \sum_{j=1}^{L} \sigma_j^x .
\end{align}
Here $\sigma_j^{\alpha}$ are the usual Pauli matrices, $B$ represents a longitudinal field which we henceforth set $B=0$, and the couplings $\J_j = \J(Q j )$ and $\h_j = \h(Q j )$ are obtained from sampling $2\pi$-periodic functions $\J(\theta)$ and $\h(\theta)$ with wave-number $Q$. Quasi-periodicity requires that the ratio of the wavelength $2\pi/Q$ to the lattice length $a=1$ is irrational: 
\begin{align}
    Q \not \in 2 \pi \mathbb{Q}.
\end{align}
Our analysis focuses on pure tone sinusoidal modulation:
\begin{align}
\J(\theta) & = \BJ + \AJ \cos(\theta +Q/2 + \phi) \nonumber\\
\h(\theta) & = \Bh + \Ah \cos(\theta+\phi+\Delta).
\label{eq:SineMod}
\end{align}
where $\BJ,\Bh,\AJ,\Ah>0$ without loss of generality. This model is depicted in Fig.~\ref{Fig:IsingChain}a.
The results for the single tone case easily generalise to generic \emph{continuous} $\J(\theta)$, $\h(\theta)$. In Tab.~\ref{tab:ModIsing} and generally, if either of $\J(\theta)$, $\h(\theta)$ has zeroes, the modulation is termed \emph{strong-QP}; whereas if either has discontinuities it is termed \emph{discontinuous-QP}.

The QP-TFIM (with $B=0$) is Ising-symmetric. That is, $H = PHP$ for $P = \prod_i \z_i$. The ground state phases are classified according to this symmetry: the paramagnetic (PM) phase is Ising symmetric, while in the ferromagnetic (FM) phase the symmetry is spontaneously broken. 

The QP-TFIM satisfies Ising duality. Under the transformation:
\begin{equation}
(\sigma_i^x \sigma_{i+1}^x, \,\sigma_i^z) \to (\tau_{i+1}^z, \, \tau_{i}^x \tau_{i+1}^x)
\label{eq:IsingDuality}
\end{equation}
the QP-TFIM with couplings $h_i$, $J_i$ maps to another QP-TFIM with couplings $h_i' = J_{i-1}$, $J_i'=h_i$. Thus any self dual points coincide with phase transitions.

\subsection{Commensurate approximation}
\label{sec:commensurate}

We may approach the limit of QP (i.e. incommensurate) modulation through a series of commensurate approximations $\Q = 2\pi p_i/q_i$, where the co-prime integers $p_i,q_i$ constitute the $i$th \emph{best rational approximation} to the irrational $Q/2\pi$. 
The best rational approximations $p/q$ to an irrational $z$ are those which minimise $|z-p/q|$ over all rationals with a denominator no larger than $q$. 
The incommensurate limit is obtained on taking $q_i \to \infty$.

As per the elementary results of Diophantine approximation~\cite{cassels1957introduction}, the best approximations $p_i/q_i$ are given by truncating the continued fraction expansion,
\begin{equation}
\frac{Q}{2 \pi} = a_0 + \frac{1}{a_1 + \frac{1}{a_2 + \frac{1}{a_3 + \ldots}}},
\label{eq:parfrac}
\end{equation}
at the $i$th level. For specificity, we focus on the Golden Ratio $Q/2\pi=\tau \equiv (1+\sqrt{5})/2$, for which the best rational approximations are $p_i/q_i = F_{i+2}/F_{i+1}$ where $F_i$ are the Fibonacci numbers. However, our results are readily generalisable to $Q/2\pi$ equal to any \emph{badly approximable number}. Badly approximable numbers are defined by the property that $\max_i a_i$ is finite.

In the commensurate approximation, the QP-TFIM~\eqref{Eq:ModelHam} is invariant under translations by $q_i$ lattice sites. The modes of the system are Bloch waves which can be calculated exactly in the infinite system limit $L \to \infty$. On length scales $\ell < q_i$, the scaling properties of correlation functions is controlled by the critical properties of QP-Ising universality class, whereas on scales $\ell \gg q_i$ the periodicity is apparent, and the scaling of correlations is correspondingly dictated by the Onsager universality class. Thus $q_i$ plays the role of a finite size cut-off to the QP-Ising transition.

\subsection{Jordan-Wigner transformation to Majorana fermions}

Using the Jordan-Wigner transformation
\begin{equation}
\begin{aligned}
    \gamma_{2i-1} &= \z_1 \cdots \z_{i-1}\x_i \\
    \gamma_{2i} &= \z_1 \cdots \z_{i-1}\y_i,
\end{aligned}
\label{eq:JWtrans}
\end{equation}
the QP-TFIM~\eqref{Eq:ModelHam} maps to a quadratic Hamiltonian (see Fig~\ref{Fig:IsingChain}b):
\begin{equation}
\begin{aligned}
H &= \frac{\I}{2} \left( \sum_{j=1}^{L-1}  \J_j \gamma_{2 j} \gamma_{2 j+1} + \sum_{j=1}^{L} \h_j \gamma_{2 j-1} \gamma_{2 j}\right) \\
& =  \frac{1}{4} \sum_{i,j=1}^{2L} \mathcal{H}_{ij} \gamma_i \gamma_j.
\label{eq:Hgam}
\end{aligned}
\end{equation}
where $\gamma_i$ are Majorana fermions satisfying $\{ \gamma_i , \gamma_j \} = 2 \delta_{ij}$. The antisymmetric-Hermitian matrix $\mathcal{H}$ has non-zero elements $\mathcal{H}_{ij}$ only for $|i-j|=1$. The eigenvalues of $\mathcal{H}$ come in $\pm$ pairs $\epsilon_\alpha = -\epsilon_\beta$, whose corresponding eigenvectors are related by complex conjugation $\psi_j^\alpha = \overline{\psi_j^\beta}$. Let $\alpha = 1 \ldots L$ label the $L$ positive eigenvalues. Define the Majorana fermions:
\begin{equation}
\begin{aligned}
\eta_{2\alpha-1} = \sqrt{2} \sum_{j=1}^{2L} \Re (\psi_j^\alpha) \gamma_j, \quad
\eta_{2\alpha} = \sqrt{2} \sum_{j=1}^{2L} \Im (\psi_j^\alpha) \gamma_j.
\end{aligned}
\label{eq:majamodes}
\end{equation}
where $\{ \eta_\alpha , \eta_\beta \} = 2 \delta_{\alpha\beta}$. Re-writing $H$ in terms of these Majorana fermions
\begin{equation}
H = \frac{\I}{2} \sum_{\alpha=1}^L \epsilon_\alpha \eta_{2\alpha - 1}\eta_{2 \alpha} = \sum_{\alpha=1}^L \epsilon_\alpha \left(c_\alpha^\dag c_\alpha - \frac12 \right)
\label{eq:HamDiag}
\end{equation}
Above, the complex fermions $c_\alpha =(\eta_{2 \alpha-1}+ \I \eta_{2 \alpha})/2$ encode the excitations of the TFIM.

\subsection{Spatial structure of excitation modes}

Transport properties, such as the thermal conductivity, are dictated by the spatial structure of excitations above the ground state.

In the clean TFIM, $\mathcal{H}$ is translationally invariant, and the $\psi_j^\alpha$ are \emph{delocalised} Bloch waves. This give rise to ballistic spreading of energy which is locally injected into the system. 
In non-interacting one-dimensional models, random modulation leads to exponentially \emph{localised} excitations $\psi_j^\alpha \sim \exp{\left(-|j-j_\mathrm{loc.}|/\LL \right)}$ each with some localisation centre $j_\mathrm{loc.}$ and localisation length $\zeta$~\cite{anderson1958absence}.

Similar localisation of all excitations is seen in the equilibrium phases of randomly modulated~\cite{pfeuty1979exact}, or strongly QP modulated Ising chains~\cite{Chandran:2017ab,crowley2018quasi}. 
At the transition, the modulation-induced localisation competes with the development of long-range order, which necessitates an extended soft mode at zero energy. 
This forces $\LL$ to diverge as a function of energy 
\begin{equation}
  1/\zeta  \xrightarrow{\epsilon \to 0} 0.
\end{equation}
In mesoscopic systems, this produces a vanishing fraction of delocalised low energy states with $\zeta \gtrsim L$. 
Certain QP-modulation leads to excitations with fractal structure~\cite{kohmoto1986quasiperiodic,kohmoto1987critical,hiramoto1988dynamics,fujiwara1989multifractal,hiramoto1992electronic,han1994critical,piechon1996anomalous}. 
Wavepackets formed from fractal modes spread sub-ballistically but without bound~\cite{ketzmerick1997determines}, so they are \emph{delocalised}.

\subsection{Scaling limit and scaling content}

At a phase transition, correlation functions become scale free~\cite{Goldenfeld:1992aa}. 
In the vicinity of the transition, single parameter scaling posits that correlation functions are controlled by a single length scale $\xi$ and time scale $\xi_t$ which both diverge at the transition:
\begin{align}
    \xi \sim \cexp{\delta}^{-\nu} \qquad \xi_t \sim \xi^z
\end{align}
Above, $\cexp{\delta} = \cexp{\log( J_i / h_i)}$ is the average deviation from the transition, and $\nu$ and $z$ are respectively the correlation length and dynamic critical exponents. Here, and throughout the manuscript, $\cexp{\cdot}$ denotes spatial averaging (averaging over the site index). 
The dynamic critical exponent also controls the long-wavelength features of the dispersion $\epsilon \sim |k|^z$ and the low energy features of the density of states $\rho(\epsilon) \sim \epsilon^{1/z-1}$.

In a homogeneous system ($\AJ = 0, \Ah=0$), the scales $\xi,\xi_t$ determine the correlations in the vicinity of the transition
\begin{equation}
 \gsexp{\x_i(t)\x_{i+r}(0)}_\mathrm{c} \sim \frac{1}{|r|^{2 \Delta_\sigma}}\mathcal{C}_{xx}\left(\frac{r}{\xi},\frac{t}{\xi_t}\right),
\end{equation}
where $\Delta_\sigma$ is the spin \emph{scaling dimension} and $\gsexp{ \cdot }_c$ denotes the connected part of the ground state correlator. 
In a spatially inhomogeneous systems,  $\gsexp{\x_i(t)\x_{i+r}(0)}_\mathrm{c}$ varies with the position $i$. 
One can define \emph{mean} and \emph{typical} correlators by taking either the spatial arithmetic-mean or the spatial geometric-mean respectively, and these may display different scaling behaviour~\cite{fisher1992random,fisher1995critical,fisher1999phase,motrunich2000infinite,crowley2018quantum}.

In this manuscript, we focus on the mean correlators, as these determine macroscopic physical quantities via linear response. These mean correlators similarly define scaling functions
\begin{equation}
\cexp{ \gsexp{\x_i(t)\x_{i+r}(0)}_\mathrm{c} } \sim \frac{1}{|r|^{2 \Delta_\sigma}}\mathcal{C}_{xx}\left(\frac{r}{\xi},\frac{q}{\xi},\frac{t}{\xi_t}\right).
\label{eq:scalingform}
\end{equation}
where we have included the dependence on the period of commensurate modulation $q$ (see Sec.~\ref{sec:commensurate}), which functions much like a finite size cut-off. 
The critical data $\mathcal{C}_{xx}, \Delta_\sigma, \nu, z$ of the inhomogeneous case may be altered from the homogeneous case.

The susceptibility $\chi$ to a longitudinal field $B$ is an example of a physical quantity controlled by a mean correlator. 
This diverges at the critical point $\chi \sim [\delta]^{-\gamma}$. Differentiating the free energy density $f$ we find
\begin{align}
    \chi & = - \left. \pdev{^2f}{B^2} \right|_{B=0} \nonumber \\
    & =   \sum_r \int_{-\beta/2}^{\beta/2} d \tau \cexp{\gsexp{ \x_i(0)\x_{i+r}(i\tau) }_\mathrm{c} }, \nonumber \\
    & \sim  \int dr \int_{-\beta/2}^{\beta/2} d \tau \frac{1}{|r|^{2 \Delta_\sigma}} \mathcal{C}_{xx}\left(\frac{r}{\xi},\frac{i \tau}{\xi_t}\right).
    \label{eq:chidev1}
\end{align}
The dependence on $\xi, \xi_t$ can be scaled out of the above integral, yielding the relation 
\begin{equation}
    \chi \sim  \xi_t \xi^{1-2\Delta_\sigma} \sim [\delta]^{-\nu(1+z-2\Delta_\sigma)} = [\delta]^{-\gamma}.
    \label{eq:chidev2}
\end{equation}
This provides a means to access susceptibility exponent $\gamma$ from the scaling of spatially averaged correlation functions. The clean TFIM is a well-known example of the Onsager universality class~\cite{onsager1944crystal} with exponents $z=1$, $\nu=1$, $\gamma = 7/4$ and $\Delta_\sigma = 1/8$.

\section{Phase diagram of the QP-TFIM}
\label{sec:phase_diag}

\begin{figure}
\begin{center}
\includegraphics[width=\columnwidth]{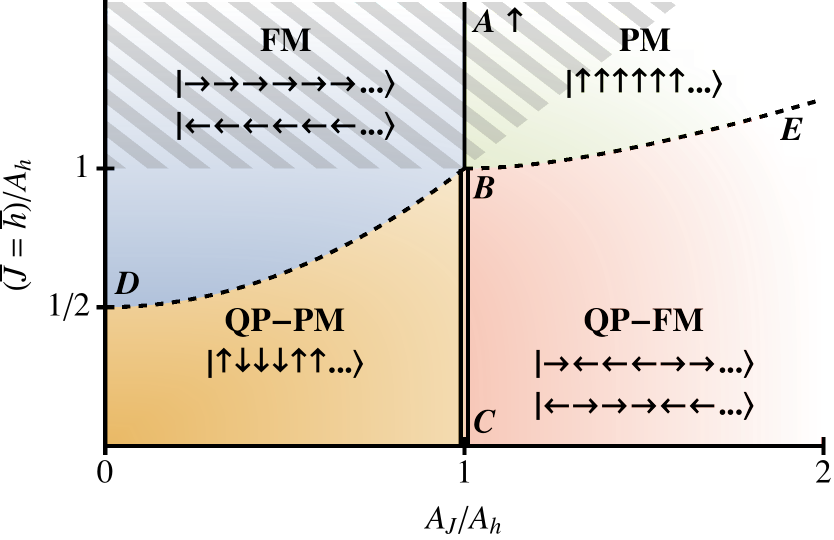}
\caption{
\emph{Phase diagram for sinusoidal couplings~\eqref{eq:SineMod}}. 
The hatched region defines the weakly modulated regime where there are no weak couplings ($J(Qi), \Gamma(Qi) > 0$ $\forall i$). In this region we find the usual gapped ferromagnetic (blue) and paramagnetic (green) phases. These are separated by a continuous transition in the clean Ising class (segment $AB$). In the strong modulation (unhatched region) excitations are localised. Within the strong modulation region we find two new modulated gapless phases: the QP-PM (yellow), and the QP-FM (red). The continuous transitions between (double line) and out of these phases (dashed lines) are in the new QP Ising class.}
\label{Fig:FigPhaseD}
\end{center}
\end{figure}

\subsection{Magnetic ordering of the phases}

We highlight an interesting slice, $\BJ = \Bh$, of the ground state phase diagram of the QP-TFIM Eq.~\eqref{Eq:ModelHam} in Fig.~\ref{Fig:FigPhaseD}.
There are four phases. When typical exchange coupling is larger than the typical field the system magnetically orders~\cite{pfeuty1979exact}. In the magnetically ordered phases if the couplings $\J_j$ are weakly modulated, $\BJ > \AJ$, the phase is the usual gapped FM phase of the clean TFIM, and neighbouring spins align. However, if the couplings $\J_j$ are strongly modulated, $\BJ < \AJ$, the system is in the gapless QP-FM phase in which spins either align or anti-align with their neighbours depending on the sign of the $\J_j$. By duality the analogous statements hold for the PM and QP-PM phases which occur when the fields are weakly ($\Bh > \Ah$) or strongly ($\Bh < \Ah$) modulated respectively.

In the FM and PM phases, the \emph{local} magnetizations (eg. $\gsexp{\sigma^x_j}$ and $\gsexp{\sigma^z_j}$) vary smoothly as the global couplings are tuned. 
In contrast in the QP-PM and QP-FM phases, these observables are  sensitive to small changes to the global couplings (i.e. $\AJ, \Ah$) as these lead to sign reversals in the local couplings. 
Nonetheless, suitably spatially averaged observables vary smoothly within these phases and can satisfy scaling near the critical boundaries.

Figure~\ref{Fig:GapMap} shows a density plot of the excitation gap across the same phase diagram as Fig.~\ref{Fig:FigPhaseD}. 
As usual, the FM and PM phases are gapped. 
The QP-FM phase is gapless because of the density of arbitrarily weak bonds across which the Ising ordering direction can locally flip at low energy and similarly for the QP-PM phase. 
We note that the QP-Ising \emph{transition} can take place between gapped QP-modulated phases when the coupling function $J(\theta), h(\theta)$ has jump discontinuities.

\subsection{QP-FM Order parameter and experimental signatures}

\begin{figure}
\begin{center}
\includegraphics[width=\columnwidth]{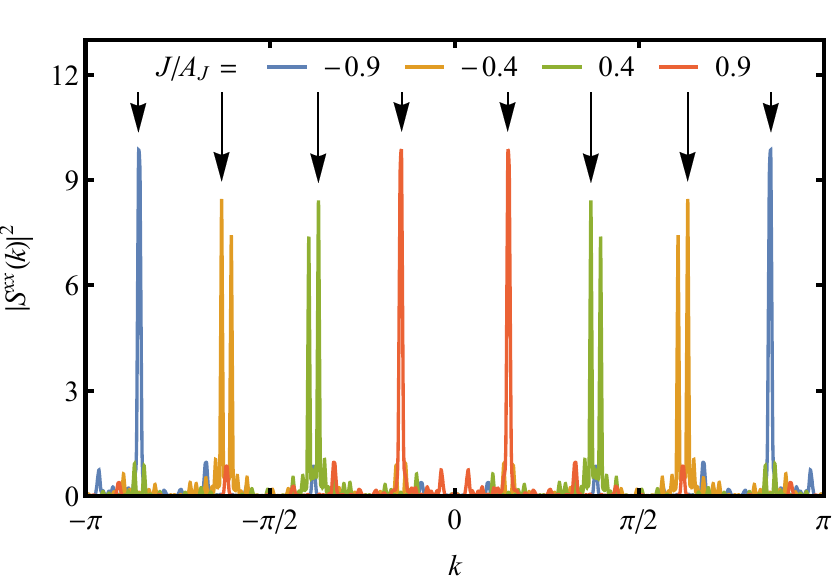}
\caption{
\emph{
Signatures in the magnetisation structure factor:
}
Upper panel: in the QP-FM limit $\AJ > \BJ \gg \Bh, \Ah$ the structure factor shows a peak at $k = \pm\kappa$ (black arrows, see Eq.~\eqref{eq:kappa}) data shown for $q = 10\,946, \Bh = \Ah = 0$ integrated over resolution scale $\delta k = 2\pi/400$. Additional subsidiary peaks are determined by the details of the QP structure of the couplings. This peak tunes between the $\kappa = 0$ (ferromagnetic order) when all the couplings are positive ($\BJ > \AJ$) and $\kappa = \pi$ (anti-ferromagnetic order) when all the couplings are negative ($\BJ < - \AJ$). Lower panel: as the transverse field strength is increased this peak persists throughout QP-FM phase before disappearing at the transition.
}
\label{Fig:StructFact}
\end{center}
\end{figure}

It is well known that the FM phase spontaneously spontaneously breaks the $\mathbb{Z}_2$ symmetry and selects one of the two degenerate ground states in which the spins are either aligned or anti-aligned to the $x$-axis. This long-range magnetic order is manifest in the non-zero value of the symmetry broken order parameter $\lim_{r \to \infty} \gsexp{\x_i \x_{i+r}} = \gsexp{\x_i}\gsexp{\x_{i+r}} \sim m^2$ where $m = \gsexp{\z_i}$ is the normalised magnetisation. 

As in the FM, in the QP-FM, the system spontaneously selects one of the two degenerate ground states related by the global spin flip. In these ground states neighbouring spins are aligned or anti-aligned according to the sign of the couplings $\J_j$. The order parameter $\gsexp{\x_j}$

The effect of sign structure is easily removed by considering the absolute value of spin-spin correlations, yielding the corresponding order parameter for the QP-FM phase $\lim_{r \to \infty} |\gsexp{\x_i \x_{i+r}}| = m^2$.

Formally $\cexp{\z_j}$ is 

For condensed matter realisations of the QP-Ising model, as with the anti-ferromagnet, experimental signatures of the order are evident in neutron scattering experiments which probe the structure factor
\begin{equation}
    S^{\alpha\beta}(k) = \sum_{r}\e^{-i k r}\cexp{\gsexp{\sigma_j^\alpha\sigma_{j+r}^\beta}}.
\end{equation}

In the QP-FM limit, $\AJ > \J \gg \h, \Ah$, the ground state correlations are given by $\gsexp{\z_j \z_{j+r}} = \gsexp{\y_j \y_{j+r}} = 0$ and $\gsexp{\x_j \x_{j+r}} = 1 (-1)$ where there are an even (odd) number of negative spin couplings in region $J_j,J_{j+1},\cdots J_{j+r-1}$. Thus as $r$ is varied the value of $\gsexp{\sigma_j^z\sigma_{j+r}^z}$ flips sign on average every $\rho^{-1}$ sites, where $\rho$ is the fraction of negative couplings. This order persists over long distances and leads to a peak in $S^{xx}(k)$ at $k = \pm\kappa$, where 
\begin{equation}
    \kappa = 2 \pi \rho = \int_0^{2\pi} d \theta \, \mathrm{sign} J(\theta) = \arccos \left(\frac{\BJ}{\AJ}\right).
    \label{eq:kappa}
\end{equation}
This peak is visible in Fig.~\ref{Fig:StructFact}, upper panel. This peak in $S^{xx}(k)$ at non-trivial $k$ is a clear signature of the quasi-periodic order and is not seen in the random Ising model which exhibits a single peak at $k=0$ when the $J_j$ are predominantly positive, and $k=\pi$ when they are predominantly negative.

For non-zero $\Bh,\Ah$, the ground state correlations are accessible only numerically, we see the peak in the structure factor persists throughout the QP-FM phases, before decreasing and disappearing at the transition (Fig.\ref{Fig:StructFact}, lower panel).

\begin{figure}[t]
\begin{center}
\includegraphics[width=\columnwidth]{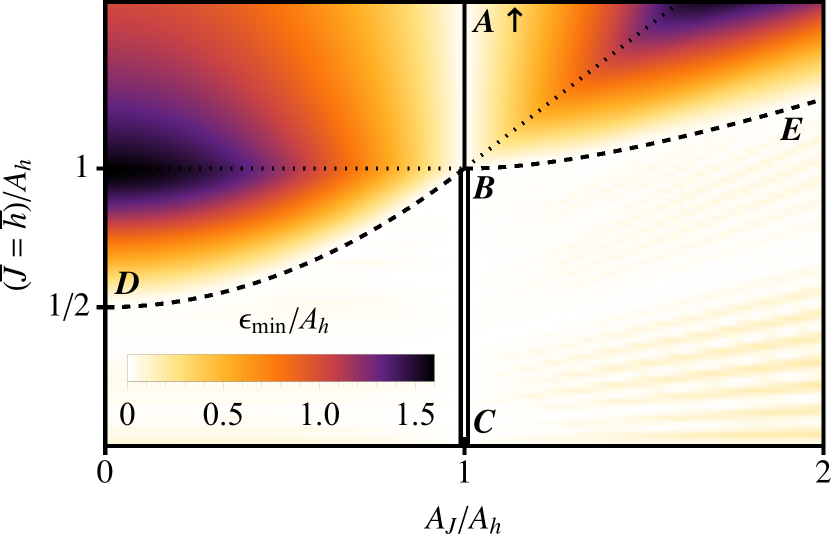}
\caption{
\emph{Spectral gap}: The solid, double-solid and dashed lines denote the phase transitions of the QP-Ising model (see Fig~\ref{Fig:FigPhaseD}). In the strongly modulated regime (below the dotted line) the excitation spectrum is localised. The value of the gap $\epsilon_{\min}/\Ah$ is denoted by colour (legend inset). The PM and FM phases (above dashed line) are gapped, whereas in the QP-FM and QP-PM phases the gap goes to zero as $q \to \infty$. Parameters: $\phi = \sqrt{3}, \Delta = \sqrt{2}, q = 144$. 
}
\label{Fig:GapMap}
\end{center}
\end{figure}

\subsection{Majorana $0$-modes and phase boundaries}
\label{sec:0mode}

The precise phase boundaries can be identified most easily by analysing the Majorana edge modes. 
In the thermodynamic limit of the FM and QP-FM phases, two of the Majorana eigen-modes $\eta_L$ and $\eta_R$ have zero energy,
\begin{equation}
[H,\eta_\mathrm{L}]=[H,\eta_\mathrm{R}]=0.
\label{eq:edgemodecomm}
\end{equation}
In the fermionic language, $\eta_L$ and $\eta_R$ are the unpaired topological edge modes of the Kitaev chain~\cite{kitaev2001unpaired}, whereas in the TFIM, they encode the two symmetry breaking ground states.

Expanding Eq.~\eqref{eq:edgemodecomm} in the basis of local fermions $\gamma_j$, the coefficients $\psi_j^{\text{L/R}}$ satisfy the two recursion relations
\begin{equation}
\psi^{\text{L/R}}_{2i+1} = \frac{\h_i}{\J_i}\psi^{\text{L/R}}_{2i-1}, \quad \psi^{\text{L/R}}_{2i-2} = \frac{\h_i}{\J_{i-1}}\psi^{\text{L/R}}_{2i}.
\label{eq:0modeshooting}
\end{equation}
Any linear combination of $\psi_j^{\text{L}}$, $\psi_j^{\text{R}}$ yields a valid zero mode. Choosing $\eta_\mathrm{L}$ and $\eta_\mathrm{R}$ to be localised at opposite ends of the chain, one finds that $\eta_\mathrm{L}$ has support only on odd sites, and $\eta_\mathrm{R}$ has support only on even sites: $\psi^{\text{L}}_{2i}=0$, $\psi^{\text{R}}_{2i+1}=0$. 

The localisation length $\LL_0$ controls the decay of the edge modes into the bulk of the chain $\psi_{2l+1}^{\text{L/R}} \sim \psi_1^{\text{L/R}} \e^{-l/\LL_0}$. Solving for the localisation length of the left mode one finds
\begin{equation}
\frac{1}{\LL_0} = \lim_{l \to \infty} \frac{1}{l} \log \left| \frac{\psi_1}{\psi_{2l+1}} \right| = \lim_{l \to \infty} \frac{1}{l} \sum_{j=1}^l \log \left| \frac{\J_j}{\h_j} \right| \equiv \cexp{\delta_j}.
\label{eq:LocLength}
\end{equation}
Here the local reduced coupling is 
\begin{equation}
\delta_j = \log |\J_j| - \log |\h_j|.
\end{equation}

At the transition out of the symmetry breaking phase, the zero modes mix into bulk modes and cease to exist. 
For the edge modes to mix with bulk modes their localisation length must diverge. This gives the condition for criticality
\begin{equation}
 \cexp{\delta_j} = 0.
 \label{eq:criticalcond}
\end{equation}
Eq.~\eqref{eq:criticalcond} corresponds to the familiar condition $\cexp{\log |\J_j|} = \cexp{\log |\h_j|}$ for the critical point of the random TFIM~\cite{pfeuty1979exact}.

As the sequence $Qj\mod 2\pi$ is equi-distributed on the interval $[0,2\pi]$, we may re-cast the sum in Eq.~\eqref{eq:criticalcond} into an integral:
\begin{equation}
\begin{aligned}
\cexp{\delta_j} &= \lim_{l \to \infty} \frac{1}{l} \sum_{j=1}^l \log \left| \frac{\J_j}{\h_j} \right| \\ 
& = \int_0^{2\pi} \frac{d \theta}{2 \pi} \log \left|\frac{\BJ + \AJ \cos (\theta+Q/2)}{\Bh + \Ah \cos (\theta+\Delta)} \right|.
\label{eq_deltaint}
\end{aligned}
\end{equation}
The zeros of this integral may be obtained analytically~\cite{Chandran:2017ab}. In the $\Bh = \BJ$ plane, this yields the phase boundaries
\begin{subequations}
\begin{align}
\Ah &= \AJ, & \label{eq:VertCrit} \\
\frac{(\BJ = \Bh)}{\Ah} &= \frac{1 + (\AJ/\Ah)^2}{2}  & \text{for} \,\,  \AJ<\Ah, \label{eq:ParaCritL}\\
\frac{(\BJ = \Bh)}{\Ah} &= \frac{ (\AJ/\Ah)^{-1} + \AJ/\Ah}{2} & \text{for} \,\,  \AJ>\Ah. \label{eq:ParaCritU} 
\end{align}
\label{eq:critlines}
\end{subequations}
These lines are shown in Fig.~\ref{Fig:FigPhaseD}. They meet at the bi-critical point $\BJ =\Bh = \AJ = \Ah$.  
Under the action of the duality transformation~\eqref{eq:IsingDuality} the line~\eqref{eq:VertCrit} is self dual, whereas~\eqref{eq:ParaCritL} and~\eqref{eq:ParaCritU} are interchanged.

Note that the phase boundaries depend only on the energetic scales $\BJ, \Bh, \AJ, \Ah$ of the model, and are independent of the wave vector $\Q$ and the phases $\phi$ and $\Delta$.

\section{Wandering of QP modulation}
\label{sec:QPscaling}

The primary effect of quasi-periodic modulation on the critical TFIM is captured by the its \emph{wandering}.
In this section, we define and analyse the wandering itself and in Sec.~\ref{sec:eqexponents} we consider the implications for the critical data.

The wandering records the variation of the reduced coupling $\delta_j$ when summed over regions of finite length $l$. The Onsager universality of the clean TFIM may persist in the presence of modulation only if a criterion due to Harris~\cite{Harris:1974aa} and Luck~\cite{Luck:1993ad,Luck:1993fu} is satisfied. 
We find generic strong continuous and discontinuous QP modulation violates this criterion, albeit much more weakly than random modulation, and thus leads to universality which is intermediate to the clean and random cases. 

\subsection{Distinct cases analysed}
\label{sec:cases}
Up to this point, our analysis has applied to the QP-TFIM irrespective of the value of the phase difference $\Delta$.
However, on the self-dual critical boundary $\AJ = \Ah$, when the value $\Delta/Q$ takes special rational values lead to fine tuned critical behaviour, distinct from the generic case.
Thus, we separate our discussion into the following cases (cf. Table~\ref{tab:Isingexponents}):
\begin{enumerate}
\setcounter{enumi}{-1}
    \item \textit{Ising}: When $\AJ<\BJ$, $\Ah<\Bh$ the modulation is \emph{weak} and irrelevant to the clean Ising transition~\cite{Luck:1993ad,Luck:1993fu}, independent of $\Delta$.
    \item \textit{QP Ising:} For strong modulation ($\AJ > \BJ$ or $\Ah > \Bh$) with generic $\Delta$ on the critical lines $DB$, $BE$, $BC$, the universality class of the transition is QP-Ising, with universal content completely determined by the wandering coefficient $w$~\cite{crowley2018quasi}.
    \item \textit{Zero-Wandering}: For strong modulation on the self-dual boundary ($BC$) with $\Delta = Q(d + 1/2)$ for $d \in \mathbb{N}$, the wandering coefficient vanishes due to fine tuning and the Harris-Luck criterion is satisfied by the clean Ising transition. However, we find that the critical data are nonetheless modified and the system behaves as if in the QP-Ising class but with a broken relationship between wandering $w$ and critical exponents. 
    \item \textit{Aubry-Andr\'{e}}: Strong modulation with $\Delta = 0$ on the self-dual boundary $BC$, the wandering coefficient is again finite and we find the equilibrium scaling content is described by the generic QP-Ising transition (Case 1).
    However, the excitations are de-localised at all energies due to an Aubry-Andr\'e type symmetry. 
\end{enumerate}

\subsection{Harris-Luck Criterion}
\label{eq:HL}

The Harris-Luck criterion concerns the behaviour of the wandering, which is defined as the sum of reduced couplings over a region of length $l$
\begin{equation}
S_{l}(j) = \sum_{i=j}^{j+l-1} \delta_i.
\label{eq:SL}
\end{equation}
This quantity characterises the local deviation from criticality over the region, $\delta_\mathrm{local}(l) = S_{l}(j)/l$. $\delta_\mathrm{local}(l)$ has mean value $\cexp{\delta_j}$ and typical fluctuations of scale $\sigma(S_l)/l$, with
\begin{equation}
\sigma(S_l) = \sqrt{\cexp{S_{l}(j)^2} - \cexp{S_{l}(j)}^2}.
\end{equation}
We decompose the local averaged reduced coupling into its mean value, and fluctuations about the mean
\begin{equation}
\delta_\mathrm{local}(l) \sim \cexp{\delta_j} + c_j\, \sigma(S_l)/ l
\end{equation}
where $c_j$ is some $O(l^0)$ number dependent on microscopic details. It is clear that $\delta_\mathrm{local}(l)$ cannot converge to its mean value in the limit of large $l$ if the fluctuations are asymptotically larger than mean. 
This imposes the consistency condition
\begin{equation}
    \lim_{l \to \infty} \sigma(S_l)/(l |\cexp{\delta_j}|) < \infty
    \label{eq:irrel_modulation}
\end{equation}
To see how this condition bounds the critical exponents, set $l$ to  $\xi$, the length-scale up to which the critical point controls the ground state correlations. 
This recasts~\eqref{eq:irrel_modulation} as the Harris-Luck criterion for the stability of the transition to spatial modulation:
\begin{equation}
\lim_{\xi \to \infty} \sigma(S_\xi)/ \xi^{1-1/\nu} < \infty
\label{eq:HLC}
\end{equation}

Random modulation provides a useful example. In this case $\sigma(S_\xi) \sim \sqrt{\xi}$ whilst in the clean TFIM $\nu=1$. These quantities violate~\eqref{eq:HLC}, indicating that in the vicinity of the transition, the fluctuations in $\cexp{\delta}$ on the length scale $\xi \sim [\delta]^{-1}$ are too large to determine the phase of the system. Random modulation is therefore a relevant perturbation to the clean Ising transition. The random Ising chain flows to an infinite-randomness critical point with $\nu = 2$~\cite{fisher1992random,fisher1995critical,fisher1999phase}, the minimal value which satisfies~\eqref{eq:HLC}.

\subsection{Case 0: Ising}

\begin{figure}
\centering
\includegraphics[width=0.48\textwidth]{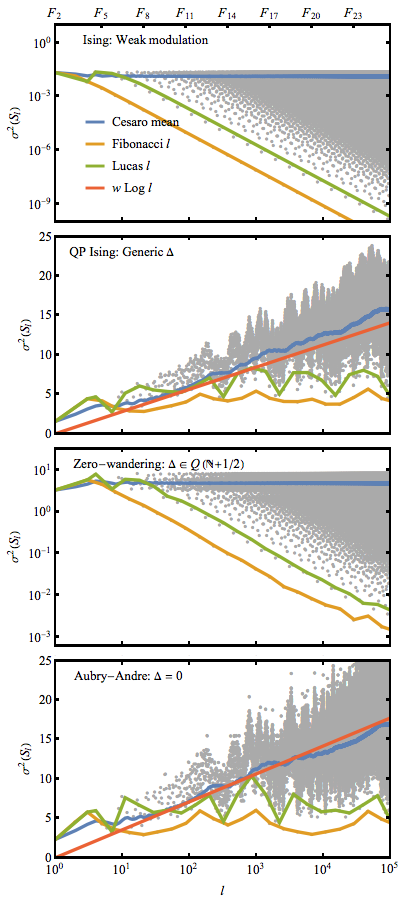}
\caption{
\emph{Logarithmic wandering of $\sigma^2(S_\ell)$:} $\sigma^2(S_\ell)$ (grey dots) has diverging infimum and supremum. The C\'{e}saro mean is shown (blue), as are sub-series with $l = $ Fibonacci (gold) and Lucas numbers (green).
\emph{Ising case}: the supremum is bounded by a constant, and the infimum decreases exponentially. 
\emph{QP Ising case}: The supremum and C\'{e}saro mean increase logarithmically. 
The analytic prediction of Sec.~\ref{sec:logwander} is shown in red.
\emph{Zero wandering case}: $\Delta \in Q(\mathbb{N}+1/2)$, the wandering has the same qualitative behaviour as Case 0. 
\emph{Aubry-Andr\'{e} case}: $\Delta = 0$ shows the same qualitative behaviour as Case 1, with slightly larger $w$. 
Parameters $(\J = \h)/(\AJ = \Ah) = 2,1/2$, and $Q/2\pi = \tau$ the Golden ratio.}
\label{fig:bball}
\end{figure}

In the QP-TFIM, we use the equivalence of spatial averages $\cexp{\cdot}$, and phase averages $\cexp{\cdot}_\phi$ to recast $\sigma^2(S_l)$ in a simple form:
\begin{equation}
\sigma^2(S_l) = \sum_{k\neq 0} |\hat\delta_k|^2 \frac{\sin^2(\Q k l/2)}{\sin^2(\Q k /2)}.
\label{eq:SigEll}
\end{equation}
Above, the Fourier coefficient $\hat\delta_k$ is defined as:
\begin{equation}
\delta(\theta) = \log \left| \frac{\J(\theta)}{\h(\theta)}\right| = \sum_k \hat\delta_k \mathrm{e}^{\mathrm{i} k \theta}.
\label{eq:delFourier}
\end{equation}
The values of $\sigma^2(S_l)$ in the weakly modulated regime ($\BJ>\AJ$, $\Bh > \Ah$) are depicted in Fig.~\ref{fig:bball} (upper panel, grey dots). We see that $\sigma^2(S_l)$ is a non monotonic function, bounded by its asymptotically separated infimum and supremum
\begin{equation}
l^{-2} \lesssim \sigma^2(S_l) \lesssim 1.
\end{equation}
Here $A_l \lesssim B_l$ is equivalent to $A_l< c B_l $ for some finite $c$ and all sufficiently large $l$. 
Certain sub-series saturate the lower scaling bound, for example in Fig.~\ref{fig:bball} (upper panel) $\sigma^2(S_l)$ scales as its infimum when $l$ is a Fibonacci (gold line) or Lucas (green line) number.

As the infimum and supremum are asymptotically separated, we characterise the scaling behaviour by the Ces\`{a}ro mean
\begin{equation}
\cexp{\sigma^2(S_l)}_{\text{Ces\`{a}ro}} = \frac{1}{l} \sum_{l' = 1}^l \sigma^2(S_{l'}).
\label{eq:SlCesaro}
\end{equation}
In the weakly modulated regime
\begin{equation}
\cexp{\sigma^2(S_l)}_{\text{Ces\`{a}ro}} \sim c
\end{equation}
for some constant $c$. 
At the clean Ising transition $\nu=1$ and the Harris-Luck criterion~\eqref{eq:HLC} is not violated either by the supremum or the Ces\'{a}ro mean. 
The clean Ising transition is therefore stable to the introduction of weak QP modulation~\cite{Luck:1993ad,Luck:1993fu}.

\subsection{Case 1: QP-Ising}
\label{sec:wanderingC1}

In the strongly modulated regime, $\sigma^2(S_l)$ is a non-monotonic function with an asymptotically separated infimum and supremum (Fig~\ref{fig:bball}, second panel, grey)
\begin{equation}
1 \lesssim \sigma^2(S_\xi) \lesssim \log l.
\end{equation}
As in the weakly modulated case, the Fibonacci (gold line) and Lucas (green line) numbers follow the infimum. 
The Ces\`{a}ro mean scales logarithmically
\begin{equation}
\cexp{\sigma^2(S_l)}_{\text{Ces\`{a}ro}} \sim w \log l
\label{eq:wdef}
\end{equation}
where  $w$ is the \emph{logarithmic wandering coefficient}.
The Harris-Luck criterion~\eqref{eq:HLC} is violated and the critical lines $BC$, and the parabolic phase boundaries $DB$ and $BE$ shown in Fig.~\ref{Fig:FigPhaseD} all have critical behaviour distinct from the clean model.

\subsubsection{Intuition for $\log$-wandering}

The log-wandering originate from the logarithmic divergence in $\delta(\theta)$~\eqref{eq:delFourier}. In a region of size $l$ running over sites site $i\leq j <i+l$, the values of the reduced coupling are set by $\delta(\theta)$ evaluated at $\theta = Q j \mod 2 \pi$. These values are sufficiently uniformly distributed over the interval $[0,2\pi]$ that we can gain intuition from considering $S_l$ as analogous to the Riemann sum
\begin{equation}
    \sum_{j=1}^l \delta(2\pi j/l) \approx l \int_0^{2 \pi} d \theta \delta(\theta) \sim O(l)
\end{equation}
Shifting the region of interest by varying $i$ moves this roughly uniformly lattice of $\theta$ values around, and induces fluctuations on this Riemann sum, these fluctuations are analogous to quantity $\sigma^2(S_l)$. When $\delta(\theta)$ is bounded and continuous, these fluctuations are $O(1)$, whereas when $\delta(\theta)$ has a logarithmic divergence, the fluctuations are dominated by how close one samples to the divergence and one finds $\sigma^2(S_l) \sim \log l$~\cite{crowley2018quasi}.

\subsubsection{The logarithmic wandering coefficient $w$}
\label{sec:logwander}

The logarithmic wandering coefficient $w$ controls the strength of the violation of the Harris-Luck criterion. 
As $w$ determines the universal content of the QP-Ising transition~\cite{crowley2018quasi}, we derive its precise value below. 
From the definition of $w$~\eqref{eq:wdef}, the Ces\`{a}ro mean~\eqref{eq:SlCesaro} and $\sigma^2(S_l)$~\eqref{eq:SigEll} 
\begin{equation}
\begin{aligned}
w &= \lim_{l \to \infty}\frac{1}{\log l}\sum_{k \neq 0} |\hat\delta_k|^2 \sum_{l'=1}^{l} \frac{\sin^2(\Q k l'/2)}{l\sin^2(\Q k /2)}.
\end{aligned}
\label{eq:wsum}
\end{equation}
For strongly modulated smooth couplings, the zeros in $\J(\theta),\h(\theta)$, imply that
\begin{equation}
\label{eq:cesarosum}
    \hat\delta_k = \frac{1}{2\pi} \int_0^{2 \pi} d \theta \, \mathrm{e}^{- i k \theta} \log \left| \frac{\J(\theta)}{\h(\theta)}\right| \sim \frac{1}{k}.
\end{equation}
The logarithmic growth of the sum~\eqref{eq:cesarosum} with $l$ is due to exponentially spaced $O(1)$ terms, which appear when the denominator $\sin^2(Q k/2)$ takes an $O(1/k^2)$ small value. 

Eq.~\eqref{eq:wsum} can be simplified. We note first that the series $f_k = k^2 |\delta_k|^2$ is a quasi-periodic in $k$ and has the following values on the different critical lines
\begin{equation}
\begin{aligned}
    f_k &\equiv  k^2 |\hat\delta_k|^2 \\ 
    &= 
    \begin{cases}
    T_k^2(\BJ/\AJ) & \text{ for $BE$ } \\
    4 \sin^2\left[(\Delta/2 - Q/4) k\right] T_k^2(\BJ/\AJ) & \text{ for $BC$} \\
    4 \sin^2\left[(\Delta/2 - Q/4) k\right] & \text{ for $B$}
    
    \end{cases}
    \label{eq:fk}
    \end{aligned}
\end{equation}
where $T_k(z) = \cos(k \arccos z)$ is the $k$th Chebyshev polynomial of the first kind. 
The properties of the line $DB$ follow by duality from $BE$. 
In all cases of~\eqref{eq:fk}, if $Q$ is rationally independent of $\Delta$, and $\arccos (\BJ/\AJ)$, then the $O(1)$ terms of the sum in~\eqref{eq:wsum} uniformly sample the values of $f_k$ and~\eqref{eq:wsum} factorises as
\begin{equation}
    w = \cexp{f}_{\text{Ces\`{a}ro}} w_Q.
    \label{eq:w_factor}
\end{equation}
The Ces\`{a}ro mean of $f_k$ can be evaluated on the various critical lines,
\begin{equation}
\begin{aligned}
\cexp{f}_{\text{Ces\`{a}ro}} & = \lim_{k \to \infty} \frac{1}{k}\sum_{k'=1}^k f_{k'} \\
& =
\begin{cases}
1/2 & \text{ for $DB$ and $BE$} \\
1 & \text{ for $BC$ } \\
2 & \text{ for $B$ } \\
\end{cases}
\end{aligned}
\label{eq:dck}
\end{equation}
The second factor,
\begin{equation}
\begin{aligned}
w_Q &= \lim_{l \to \infty}\frac{1}{\log l}\sum_{k \neq 0} \sum_{l'=1}^{l} \frac{\sin^2(\Q k l'/2)}{k^2 l\sin^2(\Q k /2)}.
\end{aligned}
\label{eq:wQsum}
\end{equation}
Refs.~\cite{crowley2018quasi,speyerprivate} showed that this limit converges to a finite value for $Q/2\pi$ equal to any badly approximable number. 
For example, in the case $Q/2\pi = \tau \equiv (1+\sqrt{5})/2$, $w_Q$ may be exactly evaluated~\cite{crowley2018quasi,speyerprivate}
\begin{equation}
w_{Q} = \frac{2 \pi^2}{15 \sqrt{5} \log \tau} = 1.22 \ldots.
\label{eq:WQ}
\end{equation}
This calculation of $w_{Q}$ is readily generalised to other quadratic numbers.
Putting it all together for $Q/2\pi = \tau$, 
\begin{align}
    w &= \begin{cases}
0.61\ldots & \text{ for $DB$ and $BE$} \\
1.22\ldots& \text{ for $BC$ } \\
2.44\ldots & \text{ for $B$ } \\
\end{cases}
\end{align}

\subsection{Case 2: Zero wandering}

For $\Delta = (Q d + Q/2) \mod 2 \pi$ with $d \in \mathbb{N}$ the wandering coefficient is zero due to an exact cancellation. 
As the exchange and field couplings are related by a lattice shift $\J_{j+d} = \h_{j}$, the sum $S_l(j)$ separates into two boundary pieces for $l > d$,
\begin{equation}
    S_l(j) = \sum_{i=j}^{j+d-1} \log | J_i | - \sum_{i=j+l-d}^{j+l-1} \log | h_i |.
\end{equation}
As a result, $\sigma(S_l) = \sigma(S_d)$ for all $l>d$ and the Harris-Luck bound~\eqref{eq:HLC} is not violated. 
Nevertheless, we will see later that this zero-wandering transition is not in the clean Ising universality class, due to the presence of small couplings.

\subsection{Case 3: Aubry-Andr\'{e}}

For $\Delta = 0$ on the line $BC$, the calculation proceeds similarly the QP-Ising case (Sec.~\ref{sec:wanderingC1}), and the wandering grows logarithmically as in Eq.~\eqref{eq:wdef}, with a slightly enhanced value of $w$.

The calculation of $w$ is distinct only in technical details. Specifically,
\begin{equation}
    |\hat\delta_k|^2 = \frac{4}{k^2} \sin^2\left( Q k/4 \right) T_k^2(\BJ/\AJ)
\end{equation}
The factor $\sin^2(Q k/4)$  is always $O(1)$ for $Q k/2\pi$ close to an odd integer, and always small for $Q k/2\pi$ close to an even integer.
The largest contribution to the sums in \eqref{eq:wsum} comes from these terms.

As the factor $\sin^2(Q k/4)$ is not self-averaging in the manner that allowed the factorisation~\eqref{eq:w_factor} we must instead absorb this term into the sum $w_Q$.

This results in the factorisation $w=\cexp{f}_{\text{Ces\`{a}ro}} w_Q$ (as before in~\eqref{eq:w_factor}) for $f_k = 2 T_k^2(\BJ/\AJ)$ (for which $\cexp{f}_{\text{Ces\`{a}ro}} = 1$), and 
\begin{align}
w_Q &= \lim_{l \to \infty}\frac{1}{\log l}\sum_{k \neq 0} \sum_{l'=1}^{l} \frac{2\sin^2(\Q k/4)\sin^2(\Q k l'/2)}{k^2 l\sin^2(\Q k /2)} \nonumber \\
& = 1.54 \ldots.
\end{align}

\section{Equilibrium Critical Exponents}
\label{sec:eqexponents}

We now discuss the consequences of the wandering analysis for the equilibrium critical exponents. 
The QP Ising and Aubry-Andr\'{e} cases have logarithmic wandering and are in the QP Ising universality class of Ref.~\cite{crowley2018quasi}. 
Remarkably, the zero-wandering transition also has modified critical data. 
We support the analysis in this section with numerical measurements.

\subsection{Correlation length exponent $\nu$}

% (4) Harris-Luck argument and the correlation length exponent
% 	— Harris-Luck argument for the instability of the clean Ising transition in terms of \nu and the wandering exponent
% 	— Conjecture that when the inequality is violated, \nu is modified to saturate the inequality
% 	— Value for \nu for three cases

We saw in section~\ref{eq:HL} that the Harris-Luck criterion sets a condition which must be satisfied for the phase transition to be stable to additional spatial modulation. 
Consistent with the finding in the randomly modulated TFIM~\cite{fisher1992random,fisher1995critical,fisher1999phase,crowley2018quantum}, we conjecture that, as in the randomly disordered case, the correlation length exponent is altered so that the Harris-Luck criterion is saturated
\begin{equation}
    \lim_{\xi \to \infty} \frac{\sigma(S_\xi)}{\xi |\cexp{\delta_j}|} = 1.
    \label{eq:HLCrit}
\end{equation}
Applying~\eqref{eq:HLCrit} to the generic QP-Ising and Aubry-Andr\'e transitions we find
\begin{equation}
    \delta \sim \sqrt{\log \xi}/\xi.
\end{equation}
That is $\nu = 1^{+}$, when $\nu$ is defined as $\delta \sim \xi^{-1/\nu}$ and ${}^+$ denotes a logarithmic correction. 
In contrast, for the zero-wandering transition, the correlation length of the clean Ising transition satisfies~\eqref{eq:HLCrit} and $\nu = 1$.

\subsection{Specific heat and dynamical exponent $z$}

\begin{figure}
\centering
\includegraphics[width=0.48\textwidth]{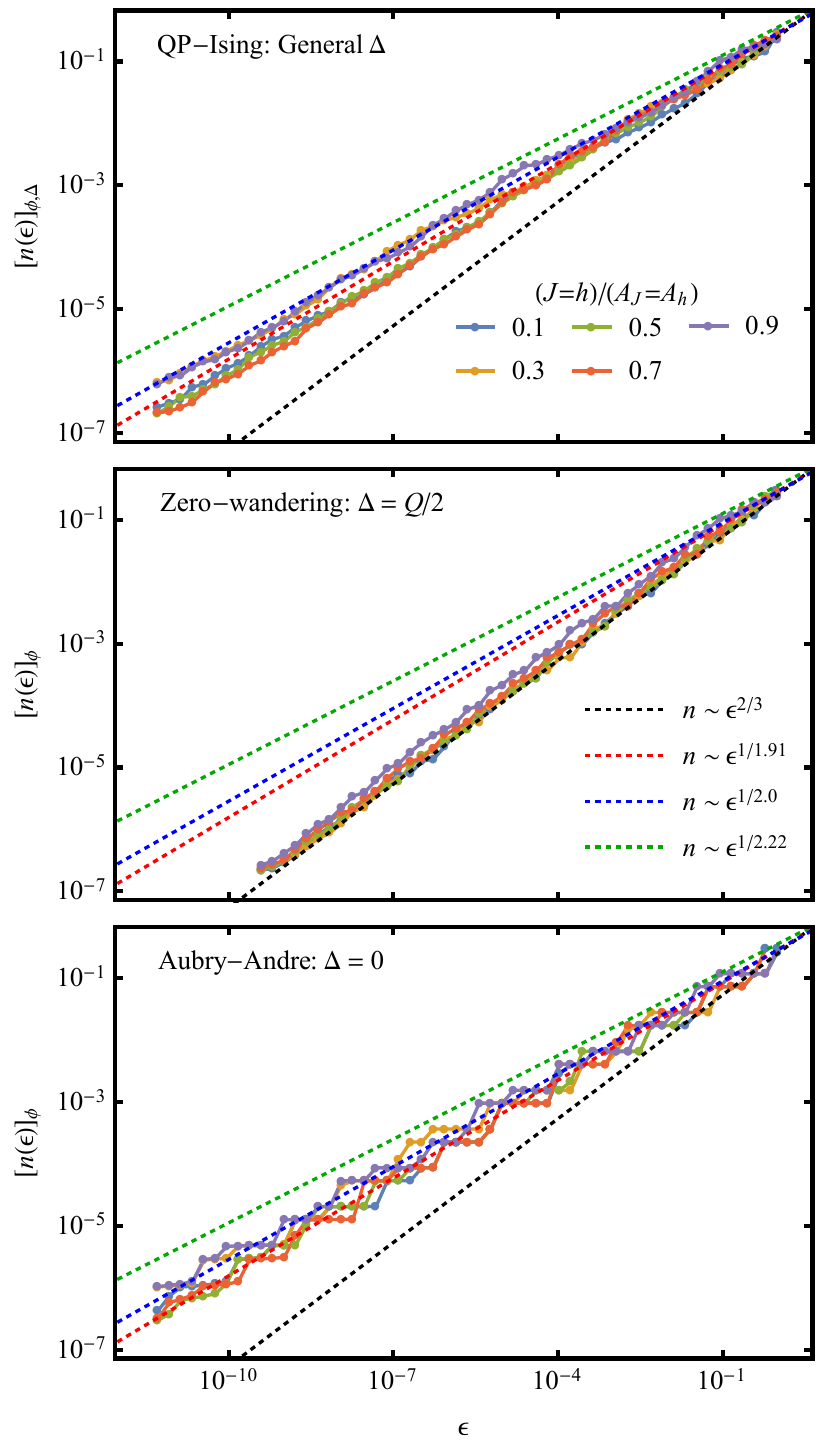}
\caption{\emph{Integrated density of states:} 
In all cases the integrated density of states scales as a power law $n(\epsilon) \sim \epsilon^{1/z}$. (QP Ising) In agreement with Ref.~\cite{crowley2018quasi}, we obtain $z \approx 1.9$ (red, dashed). This deviates from the naive estimate $z=1+w=2.22$ (green, dashed).
(Zero wandering) We obtain $z=3/2$ in agreement with calculations in Sec.~\ref{sec:zCase2} (black, dashed). 
(Aubry-Andr\'{e}) The wandering $w$ is slightly enhanced and we find an enhanced value of $z \approx 2.0$ (blue dashed). Estimates of $z$ were obtained by linear fit. Parameters: $q = 1\,346\,269$ with $n(\epsilon)$ evaluated at energies $\epsilon = \tau^{-n}$ for $n \in \mathbb{N}$.
}
\label{fig:DOS}
\end{figure}

In the vicinity of the transition, the integrated density of states obeys the following scaling
\begin{equation}
    n(\epsilon) = \int_0^{\epsilon} d \epsilon' \rho( \epsilon' ) \sim \epsilon^{1/z}
\end{equation}
We use this relationship to estimate $z$ analytically by extracting the low energy integrated density of states from a leading order approximation of the secular equation.

A macroscopic way to measure the dynamical exponent is provided by the low-temperature heat capacity 
\begin{equation}
    c = \dev{u}{T} = \dev{}{T}\int_0^\infty d \epsilon \, \epsilon \rho(\epsilon) n_\mathrm{F}(\epsilon/T).
\end{equation}
Here $n_\mathrm{F}(\epsilon/T) = (1+\mathrm{e}^{\epsilon/T})^{-1}$ is the Fermi-Dirac distribution. 
For power law density of states $\rho \sim \epsilon^{1/z-1}$,
\begin{equation}
    c = \dev{}{T} T^{1/z+1} \int_0^\infty d x \, x^{1/z} n_\mathrm{F}(x) \sim T^{1/z}
\end{equation}

\subsubsection{QP-Ising (Generic $\Delta$) and AA ($\Delta = 0$) transition}

The following calculation proceeds identically for generic $\Delta$ and $\Delta = 0$ because they both have logarithmic wandering $\cexp{\sigma^2(S_l)}_{\text{Ces\`{a}ro}} \sim w \log l$. 

Consider the excitation spectrum of the QP-TFIM. For finite period $q$, this spectrum consists of states with band index $\alpha = 1 \ldots q$, momentum $k \in [-\pi/q,\pi/q]$ and energy $\epsilon_\alpha(k)$. Let $\epsilon_\alpha^* = \max_k \epsilon_\alpha(k)$ be the highest energy of the $\alpha$th band, thus
\begin{equation}
n(\epsilon_\alpha^{*}) = \frac{\alpha}{q} \sim (\epsilon^{*}_\alpha)^{1/z}.
\end{equation}
Thus the top of the lowest band lies at an energy $\epsilon^{*}_0 \sim q^{-z}$. We note $\epsilon_0(k)$ is the root of smallest magnitude of the characteristic polynomial $\chi(\epsilon_0,k)=0$, where 
\begin{align}
    \chi(\epsilon,k) &= |\mathcal{H}(k) - \epsilon| = \prod_{\alpha = 1}^q \left( \epsilon_\alpha^2(k) - \epsilon^2 \right) = \sum_{n=1}^q \chi_{2n} \epsilon^{2n} 
    \label{eq:charpoly}
\end{align}
This allows us to estimate $\epsilon_0$ by truncating $\chi(\epsilon,k)$ to quadratic order
\begin{equation}
0=\chi(\epsilon_0) \approx \chi_2 \,  {\epsilon_0}^2 + \chi_0.
\label{eq:truncpoly}
\end{equation}
From the form of $\mathcal{H}(k)$ the coefficients $\chi_0,\chi_2$ are found to be
\begin{subequations}
\begin{align}
\chi_0 &= (-1)^q \left| \prod_i \h_i - \mathrm{e}^{- i k q} \prod_i \J_i \right|^2 \\
\chi_2 &= (-1)^{q-1} \sum_{i=0}^{q-1} \sum_{\ell=0}^{q-1}  \left( \prod_{n=i}^{i+\ell-1} \left|\J_{n}\right|^2 \prod_{n=i+\ell+1}^{i+q-1}\left|\h_{n}\right|^2 \right).
\end{align}
\end{subequations}
At the transition, $P = \left|\prod_i J_i \right| = \left|\prod_i \h_i \right|$. We therefore find
\begin{subequations}\label{eq:Chi} 
\begin{align} 
\chi_0 &= (-1)^q 2 (1 - \sigma \cos k q) P^2 \\
\chi_{2} &= q (-1)^{q-1} P^2 \sum_{l=0}^{q-1} \cexp{\frac{\mathrm{e}^{2 S_l(i)}}{|\h_{i+l}|^2}}_{i}
\label{eq:Chi2} 
\end{align}
\end{subequations}
where $\sigma = \textrm{sign}(\prod_i \J_i \h_i)$ and $S_l(i)$ is given by~\eqref{eq:SL}.
As we are interested in the maximum energy of the $0$th band, we set $ \cos(k q)=-\sigma$. 
This yields
\begin{equation}
-\frac{\chi_2}{\chi_0} =  \frac{q}{4} \sum_{l=0}^{q-1} \cexp{\frac{\mathrm{e}^{2 S_l(i)}}{|\h_{i+l}|^2}}_{i}.
\end{equation}

To make progress it is necessary to approximate further. We:
(i) neglect the correlations between the numerator and the denominator of the summand,
(ii) replace the denominator $\h_{i+\ell}$ with a single characteristic energy scale $\bar{\h}$,
and (iii) treat the $S_\ell(i)$ as Gaussian independently distributed variables with mean $\cexp{S_\ell} = 0$ and variance $\sigma^2(S_\ell) = w \log \ell$. This neglects correlations between $S_\ell(i)$ for different $i$, and non-Gaussianity of each $S_\ell(i)$. 
Making this approximation yields
\begin{align}
-\frac{\chi_2}{\chi_0} & 
 \approx \frac{q}{4 \Bh^2} \sum_{\ell=0}^{q-1}  \cexp{\mathrm{e}^{2 S_\ell(i)} }_{i} = \frac{q}{4 \Bh^2} \sum_{\ell=0}^{q-1}  \ell^{2 w} 
\sim q^{2+2 w}.
\end{align}
By~\eqref{eq:truncpoly} this estimate implies that $\epsilon_0^* \sim \sqrt{- \chi_0/\chi_2} \sim q^{-1-w}$ and hence $z \approx 1+w$. 
Using the results of Sec.~\ref{sec:logwander} for $Q/2\pi = \tau =(1+\sqrt{5})/2$, we obtain
\begin{equation}
    z \approx 1 + w = 2.22\ldots.
    \label{eq:zest}
\end{equation}
for the QP-Ising transition on the critical line $BC$. 

In Fig.~\ref{fig:DOS} this prediction is compared with numerics. The data is compared with extracted values for $z$ indicated by the dashed lines. Specifically, in each case we extract values of $z^{-1}$ by a least squares fit to the relationship
\begin{equation}
    \log [n(\epsilon)]_\phi = z^{-1} \epsilon + \mathrm{cons.}.
\end{equation}
The values of $n(\epsilon)$ are computed exactly using the method of Refs.~\cite{schmidt1957disordered,eggarter1978singular}, for $q = 1\,346\,269$. The numerics confirms the power law behaviour with an exponent $z \approx 1.9$, giving some discrepancy with the estimate~\eqref{eq:zest}. The power law behaviour of the integrated DOS $n(\epsilon) \sim \epsilon^{1/z}$ is additionally confirmed numerically for other choices of $\Q$ in the supplementary material.

\subsubsection{Zero-wandering transition ($\Delta \in Q(\mathbb{N}+1/2) \mod 2 \pi$)}
\label{sec:zCase2}

When $\Delta =  Q/2$, we have the relation $\J_j = \h_j$. 
Equation \eqref{eq:Chi} simplifies to
\begin{subequations}
\begin{align} 
\chi_0 &= (-1)^q 2 (1 - \sigma \cos k q) \left(\prod_{i} \J_i^2 \right) \\
\chi_{2} &= q (-1)^{q-1} \left( \prod_i \J_i^2  \right) \sum_{j=1}^{q} \frac{1}{J_j^2}.
\end{align}
\end{subequations}
Assuming that the sum is dominated by the minimal coupling $J_{\min} \sim 1/q$, we obtain
\begin{equation}
\epsilon_0^* \approx \sqrt{-\frac{\chi_0}{\chi_2}} = \sqrt{\frac{4}{q \sum_i \frac{1}{J_i^2}}} \sim \frac{\min |J_i|}{\sqrt{q}} \sim q^{-3/2}.
\end{equation}
Thus, $z=3/2$. 
In Fig.~\ref{fig:DOS}, we see that this agrees well with numerics.

The argument presented here is easily generalised to $\Delta = (Q d + Q/2) \mod 2 \pi$ for generic $d \in \mathbb{N}$ and predicts $z = 3/2$ provided  $q > d$. The estimate $z=3/2$ agrees well with numerics for general $d$ (data not shown).

\subsubsection{Variation of the dynamical exponent $z$ with logarithmic wandering coefficient $w$}
\label{sec:zw}

\begin{figure}
\begin{center}
\includegraphics[width=0.47\textwidth]{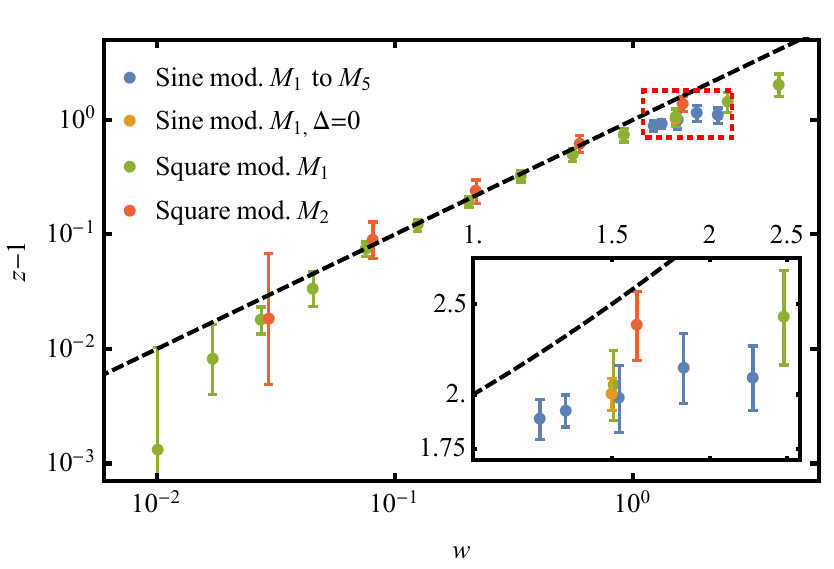}
\caption{
\emph{Variation of $z$ with $w$}: Estimates of $z$ extracted from the data shown in Fig.~\ref{fig:DOS} (sine modulation, $Q/2\pi = \tau$), Supp Mat. App. C (sine modulation, $Q/2\pi = M_n = (n+\sqrt{n^2+4})/2$ for $n=1\ldots5$) and Supp. Mat. App. D (square waves $Q/2\pi = M_1, M_2$). 
This plot confirms the approximate relationship $z = 1+w$ (black dashed) over intermediate values of $w$. 
All of the data for sine modulation fall is a small region (red-dashed box), which is enlarged in the inset.
}
\label{Fig:wzPlot}
\end{center}
\end{figure}

The QP-Ising case describes the transition for generic $\Delta$. Ref.~\cite{crowley2018quasi} conjectured that the QP-Ising critical exponents are a function of the logarithmic wandering coefficient $w$ alone. 
This conjecture is confirmed by Fig.~\ref{Fig:wzPlot}. Fig.~\ref{Fig:wzPlot} includes data for sine wave modulation~\eqref{eq:SineMod} with $Q/2\pi = M_n$, the $n$th metallic mean (blue) for $n=1\ldots 5$ (data in Fig. 11 in Supp. Mat.)
\footnote{
The metallic number $M_n \equiv \frac{n+\sqrt{n^2+4}}{2}  = n + \frac{1}{n + \frac{1}{n + \frac{1}{n + \ldots}}}$ is the number whose continued fraction expansion coefficients are all the integer $n$.
};
sine wave modulation with $\Delta = 0$, $Q/2 \pi = \tau$ (gold) (data in Fig~\ref{fig:DOS}); and square wave modulation with  $Q/2\pi = \tau \equiv  M_1$ (green) and $Q/2\pi = M_2$ (red), (data in Fig. 12 in Supp. Mat.).

The extracted values of $(w,z)$ support the conjecture that $z$ is a function of $w$ alone across a variety of QP-modulated models.
Furthermore, the analytical estimate $z \approx 1 + w$ (black dotted line) is a good approximation to $z$ over intermediate values of $w$. The  deviation at low $w$ is a finite size effect, while at large $w$, the crudeness of the approximation becomes apparent.

The square wave data in Fig.~\ref{Fig:wzPlot} is calculated from systems in which the couplings and fields take two values $\J_i \in \{ \J , \J + \AJ \}$, $\h_i \in \{ \h , \h + \Ah \}$ according to a QP sequence. This sequence is constructed in exactly the same way as the sinusoidal case: $\J_j = \J(Q j), \h_j = \h(Q j)$ but with $\J(\theta),\h(\theta)$ chosen to be $2\pi$-periodic square waves. The square wave wandering analysis (see Supp. Mat.) is a simple extension of the sinusoidal case (Sec.~\ref{sec:QPscaling}), and similarly yields logarithmic wandering. The key difference from the sinusoidal case is that the square wave logarithmic wandering coefficient $w$ has continuous parametric dependence on the ratios $\AJ/\J, \Ah/\h$ (see Supp. Mat.). Thus for square waves $w$ may be continuously tuned to zero by taking $\AJ,\Ah \to 0$ without leaving the QP-Ising universality class. 
The ability to continuously tune $w$ allows a more extensive exploration of the relationship $z(w)$.

The square wave wandering analysis generalises~\emph{mutatis mutandis} to the case of any QP sequence in which the couplings take values drawn from a finite \emph{alphabet} $J_j \in \{ \J_a, \J_b, \J_c \ldots \}$, $h_j \in \{ \h_a, \h_b, \h_c \ldots \}$. Thus generically such sequences have logarithmic wandering. However, we note that there are fine tuned sequences which also have no wandering (see Supp. Mat. App. D1b)~\cite{doria1988quasiperiodicity,igloi1988quantum,ceccatto1989quasiperiodic,kolavr1989attractors,benza1989quantum,benza1990phase,Luck:1993ad,grimm1996aperiodic,hermisson1997aperiodic,igloi1997exact,igloi1998random,hermisson1998surface}, analogous to the fine tuned Zero-wandering case. The methodology we present allows the study of TFIMs modulated by generic QP sequences whereas previous analyses have been restricted to special sequences which satisfy an inflation rule~\cite{doria1988quasiperiodicity,igloi1988quantum,tracy1988universality,ceccatto1989quasiperiodic,kolavr1989attractors,benza1989quantum,benza1990phase,lin1992phase,Luck:1993ad,turban1994surface,grimm1996aperiodic,hermisson1997aperiodic,igloi1997exact,igloi1998random,hermisson1998surface,oliveira2012strong,yessen2014properties}.

\subsection{Magnetic susceptibility and the scaling dimension $\Delta_\sigma$}

\begin{figure}
\centering
%\includegraphics[width=0.48\textwidth]{Fig_XX_Temp_1.png}
%includegraphics[width=0.48\textwidth]{Fig_XX_Temp_2.png}
%\includegraphics[width=0.48\textwidth]{Fig_XX_Temp_3.png}
\includegraphics[width=0.48\textwidth]{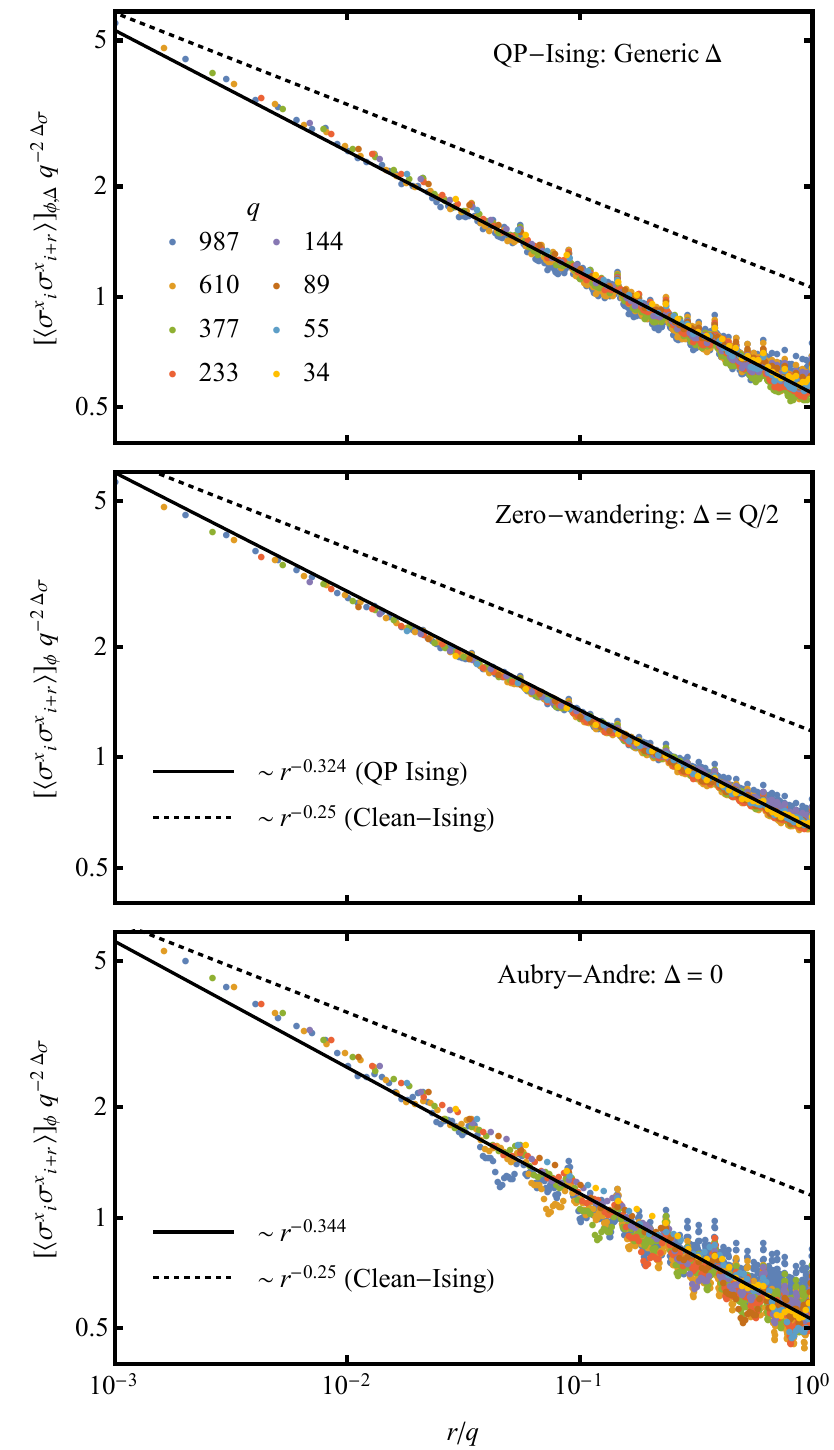}

\caption{\emph{Spin-spin correlations:} Spin-spin correlations $\gsexp{\x_i\x_{i+r}} q^{-2\Delta_\sigma}$ vs $r/q$ collapses for a range of $q$ for $\Delta_\sigma = 0.16$ (QP-Ising and Zero wandering cases) and $\Delta_\sigma = 0.17$ (Aubry-Andr\'{e} case). 
The solid line shows pure power law decay $\sim r^{-2 \Delta_\sigma}$.  This shows good fit for the QP-Ising and Zero wandering cases, indicating a simple scaling function, whereas the Aubry-Andr\'{e} case shows short range deviation from this form. The clean Ising decay with $\sim r^{-1/4}$ is shown (dotted) for comparison. 
}
\label{fig:XXCorrs}
\end{figure}

We turn to the value of the scaling dimension $\Delta_\sigma$ in the different cases. In terms of macroscopic properties of the system, $\Delta_\sigma$ controls the divergence with $\delta$ of the  magnetic susceptibility to a longitudinal field $\chi = \frac{\partial m}{\partial B}|_{B=0}$. 
Near the critical point $\chi \sim [\delta]^{-\gamma}$ with an exponent 
\begin{equation}
    \gamma = \nu (1+z-2\Delta_\sigma)
    \label{eq:scalingrel}
\end{equation}
The relationship~\eqref{eq:scalingrel} is Fishers scaling law in $d = z+1$ dimensions. The relation follows directly from the free energy in Eqs.~\eqref{eq:chidev1} and~\eqref{eq:chidev2}.

We extract $\Delta_\sigma$ by minimising the mean deviation of $\gsexp{\x_i\x_{i+r}} r^{2\Delta_\sigma}$ over $r = 1 \ldots q$ for $q = 21$ to $987$. 
We also verify data collapse for the extracted values of $\Delta_\sigma$ in Fig.~\ref{fig:XXCorrs}. In the QP-Ising and Zero wandering cases we find $\Delta_\sigma \approx 0.16$, consistent with Ref.~\cite{crowley2018quasi} which studied only the QP-Ising case. 
The Aubry-Andr\'{e} case has slightly enhanced wandering coefficient as compared to the QP-Ising value and consequently a slightly enhanced scaling dimension $\Delta_\sigma \approx 0.17$. The extracted values of $\Delta_\sigma$ and $z$ yield the values $\gamma = 2.6^+, 2.2 \text{ and } 2.7^+$ in the QP-Ising, Zero-wandering and Aubry-Andr\'{e} cases respectively. These are are larger than the Onsager value of $\gamma = 1.75$ at the clean Ising transition (see Tab.~\ref{tab:Isingexponents}).

\section{Localisation of excitations}
\label{sec:excitations}

On the critical boundary, the TFIM  possesses an extended zero energy  mode. 
The zero mode is described by Eq.~\eqref{eq:0modeshooting}, which has infinite localisation length at the critical point (cf. Eq.~\eqref{eq:LocLength})
\begin{align}
    \frac{1}{\zeta(0)} = [\delta_j] = 0
\end{align}
In the unmodulated TFIM, the zero mode is uniform; it arises as the zero energy and momentum limit of linearly dispersing fermionic low energy excitations. 
With QP modulation, the mode need not be spatially uniform. Nevertheless, it cannot be localised.
There are accordingly several scenarios for the structure of the low energy excitations:
\begin{itemize}
    \item The modes may remain ballistic, in which case the total bandwidth $W$ of the Bloch bands remains finite as the incommensurate limit is taken. The wavefunctions are uniformly extended with small spatial fluctuations.
    \item The modes may become multifractal; $W$ vanishes as a non-trivial power law in the incommensurate limit, but the finite energy inverse localisation length remains zero $1/\zeta(\epsilon) = 0$.
    \item The finite energy modes may localise, so long as the localisation length diverges as $\epsilon \to 0$:
    \begin{align}
    \zeta(\epsilon)  \sim \epsilon^{-1/z_{\mathrm{L}}}.
    \label{eq:LLE}
    \end{align}
    In this case, the total bandwidth $W$ in any small energy window at finite energy $\epsilon$ decays exponentially with $\exp(-q/\zeta(\epsilon))$.
\end{itemize}

We find that all three of these scenarios are realised. 
When the QP modulation is irrelevant to the clean Ising transition (Ising case), the low energy excitations are ballistic.
With strong modulation,  the excitations generically localise with a localisation exponent $z_{\mathrm{L}} = z$ which coincides with that extracted from the equilibrium density of states (QP-Ising and Zero-wandering cases).
This agrees with the behaviour found in previous QP~\cite{crowley2018quasi,Chandran:2017ab} and random~\cite{fisher1992random,fisher1995critical,fisher1999phase,crowley2018quantum} Ising chains.
On the other hand, in the Aubry-Andr\'{e} case, the model possesses enhanced Aubry-Andr\'e-type symmetry which requires that the localisation length be energy independent -- since it must be infinite at $\epsilon = 0$, \emph{none} of the excitation modes can localise.
In this case, the dynamical exponents decouple in the sense that $z$ remains non-trivial while $z_{\mathrm{L}}$ is not defined. 

In the following, we first provide an elementary upper bound on the total bandwidth $W$ of the TFIM in terms of the couplings in the chain and then use that as a tool to investigate excitations in each of the cases.

\subsection{Bandwidth bounds}
\label{sec:eq_ex}

In the strongly modulated regime, $\AJ > \BJ$ or $\Ah > \Bh$, there are arbitrarily small couplings in the chain. 
These small couplings force $W\to 0$ as $q\to \infty$. 
Thus, the transition in the strongly modulated regime cannot support ballistic excitations.

At finite $q$, the spectrum contains $q$ bands with energies $\epsilon_\alpha(k)$ for $\alpha = 1 \ldots q$ and Bloch momenta $k \in [-\pi/q,\pi/q]$. 
If there is a finite density of ballistic modes, then the mean (absolute) group velocity $\bar{v}$ is finite. 
Explicitly,
\begin{equation}
    \bar{v} = \cexp{ |\partial_k \epsilon_\alpha(k)| }_{k,\alpha} = \frac{1}{q} \sum_\alpha  \int_{-\pi/q}^{\pi/q} \frac{d k}{2 \pi/q} |\partial_k \epsilon_\alpha(k)|.
\end{equation}
Since the Bloch bands $\epsilon_\alpha(k)$ have only two turning points as a function of $k$ (see Supp. Mat.), 
\begin{equation}
    W = \pi \bar{v}
    \label{eq:BW_groupV}
\end{equation}
where $W = \sum_\alpha W_\alpha$ and $W_\alpha = \max_k \epsilon_\alpha(k) - \min_k \epsilon_\alpha(k)$ is the width of the $\alpha$th band. 

In the supplemental material, we prove the elementary result that the smallest coupling bounds the total bandwidth:
\begin{equation}
    W \leq 2 \pi \min_i (|J_i|,|h_i|)
    \label{eq:BW_bound}
\end{equation}
Here, the minimum runs over the couplings in the period $q$.
Since the smallest coupling in the strong modulation regime is typically $1/q$, we find
\begin{align}
\label{eq:bandwidthtozero}
    W \lesssim q^{-1} \to 0
\end{align}

Outside of the hatched region in Fig.~\ref{Fig:FigPhaseD}, Eq.~\eqref{eq:bandwidthtozero} proves the density of ballistically propagating excitations at any energy vanishes. 
The inequality is not strong enough to distinguish localisation from multifractality.
Numerically, we observe that all excitations are exponentially localised away from the phase boundaries.
The behavior on the critical line is more complicated and discussed case by case below.

\subsection{Ising case: Ballistic excitations}
\label{sub:ballistic}

For weak amplitude modulation ($\AJ < \BJ$ and $\Ah < \Bh$), all of the couplings in Eq.~\eqref{Eq:ModelHam} are finitely bounded away from zero and the bandwidth bound Eq.~\eqref{eq:BW_bound} is finite.
We find numerically that the critical excitations up to a finite mobility edge propagate ballistically as in the clean Ising model. 
This is consistent with the irrelevance of weak quasi-periodic modulation at the clean Ising critical point.

\subsection{QP-Ising and Zero wandering cases: localised excitations}

\begin{figure}
\centering
\includegraphics[width=0.48\textwidth]{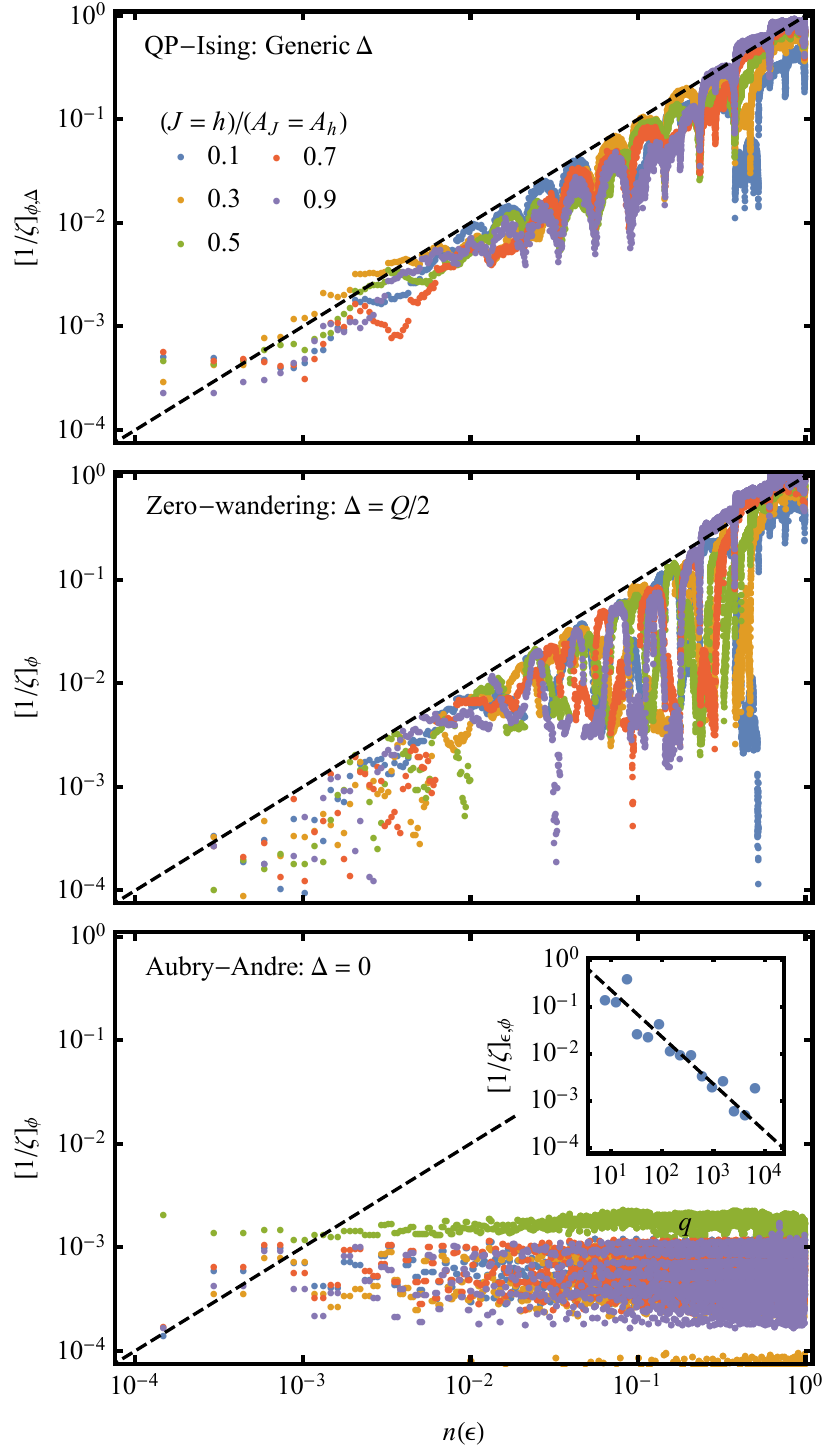}

\caption{\emph{Localisation length $\zeta(\epsilon)$:} 
The inverse localisation length $\cexp{1/\zeta(\epsilon)}$ versus the integrated density of states $n(\epsilon)$ at several points on the $BC$ critical line. The dashed line indicates the relationship $n(\epsilon) \sim \cexp{1/\zeta(\epsilon)}$. 
In the QP-Ising and Zero-wandering cases, the inverse localisation length is bounded by an envelope $\sim \epsilon^{1/z} \sim n(\epsilon)$. 
In the Aubry-Andr\'{e} case, the localisation length is $1/\zeta(\epsilon)=0$ for all $\epsilon$ as confirmed by the scaling $\cexp{1/\zeta(\epsilon)}_\epsilon \sim 1/q$ (inset). 
Parameters: $q = 6\,765$.
}
\label{fig:Locs}
\end{figure}

Numerically, the arbitrarily weak couplings in the strongly modulated regime are sufficient to localise the finite energy excitations (QP-Ising and Zero-wandering cases).
The data in Fig.~\ref{fig:Locs} confirms that the relationship
\begin{align}
    1/\LL(\epsilon) \sim n(\epsilon)
\end{align}
holds for the envelope of the inverse localisation length data and thus that $z_{\mathrm{L}} = z$. 
The visible substructure in the data is controlled by the fractal properties of the spectra and states of QP models and we do not investigate it further here.

The data for $1/\LL(\epsilon)$ is extracted from a least squares fit to the relationship
\begin{equation}
\log \left[ \left|  \psi_i^\alpha \bar\psi_{i+r}^\alpha\right| \right] = - \frac{r}{\LL(\epsilon_\alpha)}  + \textrm{const}
\end{equation}
where $\psi_i^\alpha$ is the eigenmode of $\mathcal{H}$ at energy $\epsilon_\alpha$. For Fig.~\ref{fig:Locs} we have further used that $n(\epsilon_\alpha) = \alpha/q$.

\subsection{Aubry-Andr\'{e} case: Multifractal excitations}

At the special point $\Delta = 0$, the critical delocalisation $1/\LL(0)=0$ extends to the whole spectrum $1/\LL(\epsilon)=0$.

This is enforced by a special duality which generalises the well known Aubry-Andre duality~\cite{Harper:1955kl,Azbel:1979aa,Aubry:1980aa,Hofstadter:1976aa}. The Aubry-Andre model is dual to itself under the Fourier transform. 
Many properties follow from this duality.
For example, if $\mathcal{H}$ has finite bandwidth, and hence extended modes, then its dual model has a pure point spectrum and localised modes,  and \textit{vice versa}~\cite{han1994critical,thouless1983bandwidths}. 
A corresponding duality which applies to a wider class of single particle quasi-periodic models is obtained if one considers the class of 1D short range hopping models which are dual to 1D short range hopping models~\cite{han1994critical,Chandran:2017ab}.

Consider a single particle Hamiltonian of the form
\begin{equation}
    \mathcal{H} = \sum_{j=-\infty}^\infty \sum_{a=-\infty}^\infty t_a(Qj/2) \ket{j+a}\bra{j} 
\end{equation}
where the QP modulated $a$-site hops and on-site potentials are set by the $2\pi$ periodic functions $t_a(\theta)$ and $t_0(\theta)$, respectively. Hermiticity, $\mathcal{H} = \mathcal{H}^\dag$, requires that $t_{-a}(\theta) = t_a^*(\theta-Q a/2)$. The unitary $\mathcal{V} = \frac{1}{\sqrt{\mathcal{N}}} \sum_{nm}\e^{i Q nm/2} \ket{n}\bra{m}$ Fourier transforms $\mathcal{H}$ to a dual model of the same class
\begin{equation}
    \tilde{\mathcal{H}} = \mathcal{V}\mathcal{H}\mathcal{V}^\dag  = \sum_{j=-\infty}^\infty \sum_{a=-\infty}^\infty \tilde{t}_a(Qj/2) \ket{j+a}\bra{j}
\end{equation}
where the dual hops are defined by
\begin{equation}
\tilde{t}_a(\theta) = \sum_{b=-\infty}^{\infty} \int_{-\pi}^{\pi} \frac{d \theta'}{2 \pi} \e^{i (b \theta - a \theta')} t_{b}^*(\theta')  .
\end{equation}
As the high order Fourier components of $t_a(\theta)$ contribute to long range hops in the dual basis, generic modulated nearest-neighbor hopping models are dual to models with long-range hopping.
However, for special models the hopping is local in both bases. 

On the vertical critical line at $\Delta = 0$ the single particle Hamiltonian $\mathcal{H}$ in Eq.~\eqref{eq:Hgam}, and its corresponding dual $\tilde{\mathcal{H}}$ are nearest neighbour hopping models. $\mathcal{H}$ is a tridiagonal matrix with on-site potentials and nearest neighbour hops set by,
\begin{equation}
    \begin{aligned}
        t_0(Qj/2) &=0 \\
        t_1(Qj/2) &= i \e^{i k/2}\left[\BJ + \AJ \cos\left(Q j /2 +\phi \right)\right]
    \end{aligned} 
\end{equation}
whilst $\tilde{\mathcal{H}}$ has corresponding elements,
\begin{equation}
    \begin{aligned}
       \tilde{t}_0(Qj/2) &= 2\BJ \sin(Qj/2 - k/2) \\
    \tilde{t}_1(Qj/2) &= \e^{i (\phi-Q/4)} \AJ \sin(Qj/2 + Q/4 - k/2).
    \end{aligned}
\end{equation}
The amplitude of all longer range ($a>1$) hops vanishes, $t_a ,\tilde{t}_a = 0$~\footnote{The Aubry-Andre model, and the models of Ref.~\cite{gopalakrishnan2017self} are other examples of self-dual tridiagonal models.}.

A relation due to Thouless~\cite{thouless1972relation} states that for tridiagonal model $\mathcal{H}$, the localisation length $\LL$ and density of states are related by
\begin{equation}
    \frac{1}{\LL(\epsilon)} = \int d \epsilon' \rho(\epsilon') \log |\epsilon-\epsilon'| - \cexp{\log |t_{1}(\theta)| }_\theta.
    \label{eq:thouless}
\end{equation}
As $\mathcal{H}$ and $\tilde{\mathcal{H}}$ are unitarily related, they have the same density of states. 
Applying~\eqref{eq:thouless} to both $\mathcal{H}$ and $\tilde{\mathcal{H}}$, we find that the difference of the inverse localisation lengths,
\begin{equation}
    \frac{1}{\LL(\epsilon)} - \frac{1}{\tilde{\LL}(\epsilon)}  = \cexp{\log |t_{1}(\theta)| }_\theta - \cexp{\log|\tilde{t}_{1}(\theta)| }_\theta ,
    \label{eq:LLRelation}
\end{equation}
is set by an energy independent constant. It is further known that lattice model wave-functions cannot be localised in both real and reciprocal space~\cite{han1994critical,thouless1983bandwidths}, i.e. $1/\LL(\epsilon)>0$ implies $1/\tilde{\LL}(\epsilon)=0$ and \textit{vice versa}. Thus critical delocalistion $1/\LL(0)=0$ implies the RHS of~\eqref{eq:LLRelation} is non-positive: i.e.  $\cexp{\log |t_{1}(\theta)| }_\theta \leq \cexp{\log|\tilde{t}_{1}(\theta)| }_\theta$. Hence $1/\LL(\epsilon)=0$ for all $\epsilon$.

The critical delocalisation of the excitations at all energies is verified in Fig.~\ref{fig:Locs}. The numerically extracted $1/\zeta(\epsilon)$ are found to be independent of energy (up to finite size fluctuations) and tend to zero as $q \to \infty$ (Fig./~\ref{fig:Locs}, lower panel, inset).

\subsection{Dynamics of wavepackets}
\label{sec:dynamics}

\begin{figure*}
\centering
\includegraphics[width=0.95\textwidth]{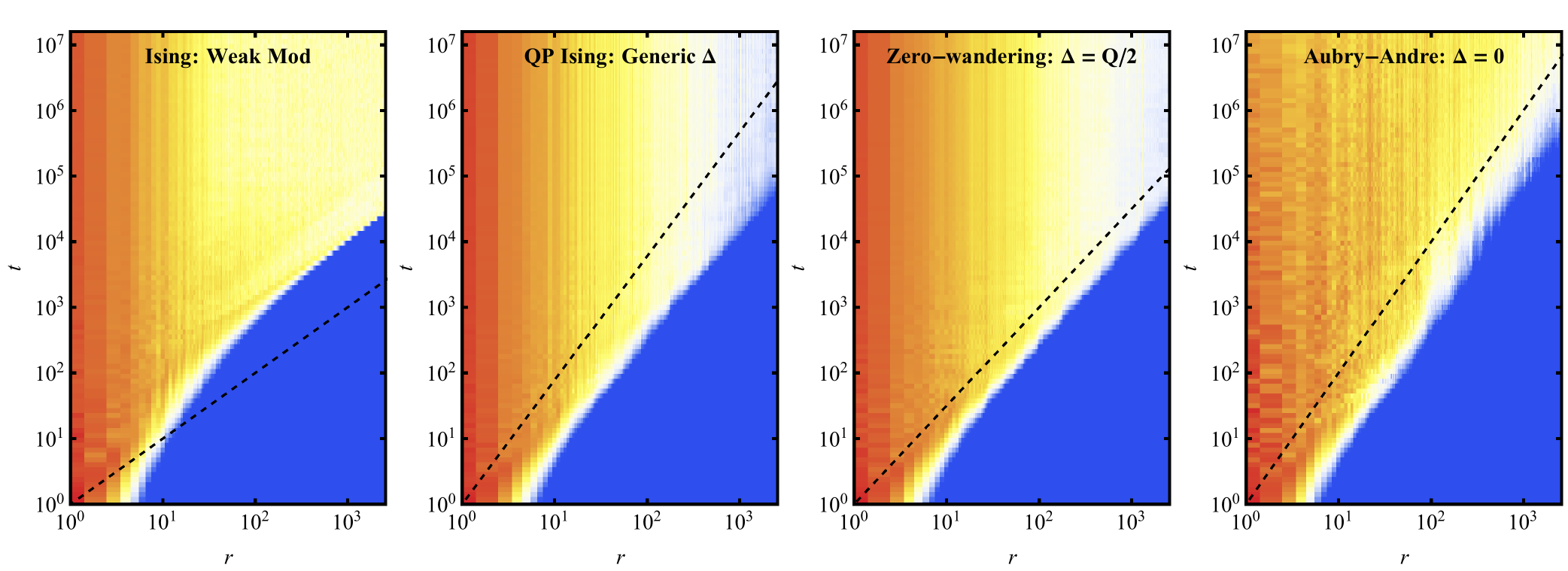}

\caption{\emph{Wavepacket spreading:} Density plots of $\log \bar{P}(r,t)$ are shown for the four cases of Sec.~\ref{sec:cases}, in each case, and $r = t^{1/z}$ are plotted as guides (dashed) using the respective values $z=1,1.89,1.5,2$ consistent with Fig.~\ref{fig:DOS}.
For the Ising case (weak modulation, left), a ballistic front $r \sim t$ is evident. This carries the weight in the delocalised low energy excitations.
The localisation of high energy excitations above a cut-off appears as weight trapped near $r=0$.
For the QP-Ising case (centre-left) and Zero-wandering case (centre-right) all but a vanishing fraction of excitations are localised. 
At distance $r$ the fraction of excitations with localisation length $\LL > r$ participate in the wave-front. This fraction vanishes as $r$ increases and $P(r,t) \sim r^{-2}$ at large $t$.
For the Aubry-Andr\'{e} case (right) all excitations are delocalised, and a diffusive wave-front is observed. The saturation to a limiting form $P(r,t) \sim r^{-\beta}$ at late times is a finite size effect (see main text and Fig.~\ref{fig:Entropy}).
Parameters: $(\BJ = \Bh)/(\AJ = \Ah) = 1.05 \text{ (left)}, 0.5 \text{ (otherwise)}$, $q = 2\,584$.
}
\label{Fig:WavePacks}
\end{figure*}

The localisation properties of the modes can also be seen in the asymptotic spreading of wavepackets. 
The spreading of a fermionic wavepacket created at site $i$ is captured by the time-evolved Majorana operators expanded in the initial basis
\begin{equation}
    \mathrm{e}^{- i H t} \gamma_i \mathrm{e}^{i H t} = \mathcal{U}_{ij}(t) \gamma_j
\end{equation}
where
\begin{equation}
    \mathcal{U}_{ij}(t) = \left( \mathrm{e}^{-i \mathcal{H} t} \right)_{ij}= 2 \Re  \sum_{\alpha=1}^q \bar{\psi}_i^\alpha \psi_{j}^\alpha \mathrm{e}^{- i \epsilon_\alpha t}.
\end{equation}
The probability of a transition from site $i$ to $j$ in time $t$ is $P_{ij}(t) = | \mathcal{U}_{ij}(t) |^2$, and we denote its spatial average by
\begin{equation}
    \bar{P}(r,t) = \cexp{P_{i,i+r}(t)}_i = \cexp{\left|\mathcal{U}_{i,i+r}(t)\right|^2}_i.
\end{equation}
This can be used as a proxy for a broad class of dynamical correlation functions $\langle \mathcal{O}_{i+r}(t) \mathcal{O}_i(0) \rangle$ as the action of any local parity-symmetric observable (ie. not involving Jordan-Wigner strings) is simply to create or destroy local Majorana excitations.

Density plots of $\log \bar{P}(r,t)$ are shown in Fig.~\ref{Fig:WavePacks}. In each case we see the wave-packet spreading to be consistent with $r \sim t^{1/z}$ spreading of excitations (black dashed lines).

In the Ising case of weak modulation, excitations below a finite mobility edge are delocalised and ballistic. 
The delocalized excitations spread without bound, forming a clearly visible ballistically propagating wavefront ($z=1$) (Fig~\ref{Fig:WavePacks}, left). 
The excitations above the mobility edge leave behind the localised remnant in the vicinity of $r=0$ (red vertical stripes).

In the QP-Ising and Zero-wandering cases, all excitations are localized, but with a diverging localization length as $\epsilon \to 0$. 
The wavefront propagates sub-ballistically to infinity with non-trivial exponent $z$ (dashed line), but the weight at the front decays asymptotically with $t$. 
More precisely, at a distance $r$ only excitation modes with a localisation length $\zeta(\epsilon) > r$ can participate in the wavefront. 
Thus, the weight decays with a power law and $\bar{P}(r,t)$ saturates to a form $\lim_{t \to \infty }\bar{P}(r,t) \sim r^{-2}$. This is seen in (Fig~\ref{Fig:WavePacks}, centre panels). The limiting form is obtained as
\begin{equation}
\begin{aligned}
   \lim_{t \to \infty} P_{ij}(t) &\sim \lim_{T \to \infty} \frac{1}{T} \int_0^T d t \, P_{ij}(t) \\
    & = 2  \sum_\alpha \left|\psi_i^\alpha \right|^2\left|\psi_{j}^\alpha \right|^2 \\
    & \sim \int d r_0 \int d \epsilon \, \epsilon^{3/z-1} \mathrm{e}^{- 2 c \epsilon^{1/z} (|r_0-i|+|r_0-j|)} \\
    & \sim \frac1{|i-j|^2}
\end{aligned}
\end{equation}
where we have used the ansatz $|\psi_r^\alpha|^2 \sim \mathrm{e}^{- 2 |r-r_0|/\LL(\epsilon_\alpha)}/\LL(\epsilon_\alpha)$, with localisation length $1/\LL(\epsilon) = c \epsilon^{1/z}$, localisation centres $r_0$ uniformly distributed over the sample, and density of state $\rho(\epsilon) \sim \epsilon^{1/z-1}$.

 \begin{figure}
\centering
\includegraphics[width=0.48\textwidth]{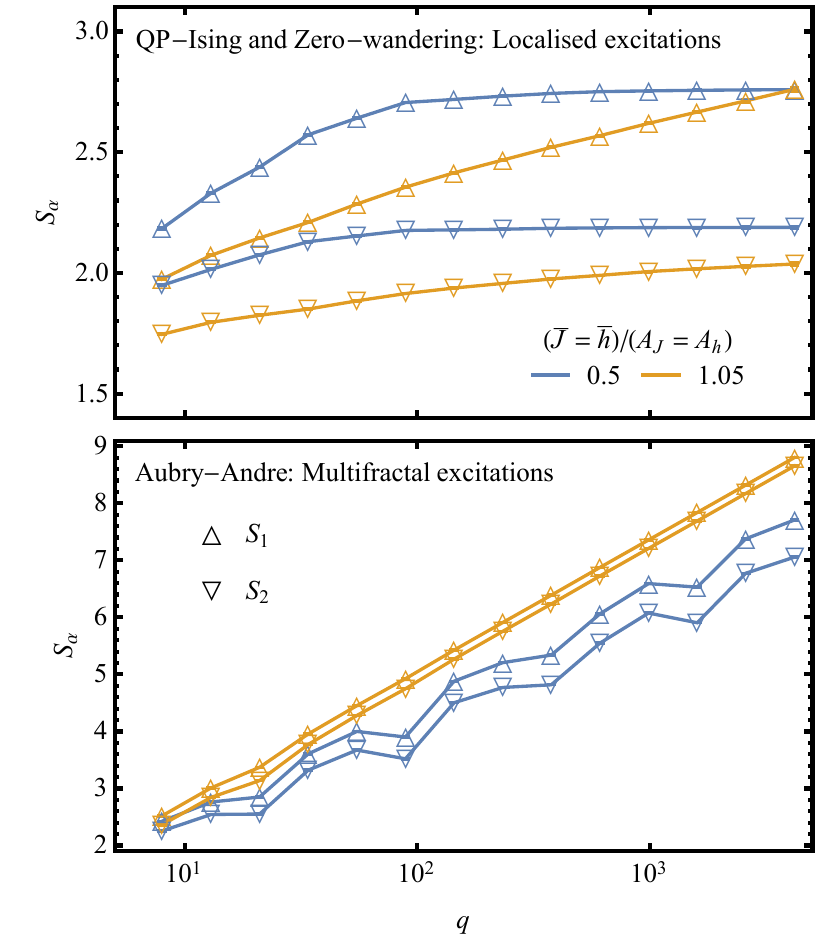}

\caption{\emph{Entropy of $P_{i,i+r}$ at finite size saturation:} the saturation of $S_1$ ($S_2$) with increasing $q$ indicate that all (some) of the weight of $P_{i,i+r}$ does not spread. Data is from the lower critical line $BC$ (blue) and the upper critical line $AB$ (gold). (QP-Ising and Zero-wandering cases) $S_1$ and $S_2$ saturate indicating bounded spreading of correlations on the lower critical line $BC$. On the upper critical line $AB$, $S_1$ is unbounded but $S_2$ saturates, due to the presence of localised and diffusive modes. (Aubry-Andr\'{e} case) Neither $S_1$ or $S_2$ saturates, indicating unbounded operator of $P_{i,i+r}$, due to the fully delocalised spectrum. Parameters: $(\BJ = \Bh)/(\AJ = \Ah) = 0.5\text{ (blue) }, 1.05 \text{ (gold) }$, $\Delta = Q/2$ (upper plot) $\Delta = 0$ (lower plot).
}
\label{fig:Entropy}
\end{figure}

In the Aubry-Andr\'{e} case the excitations are delocalised and spread asymptotically without bound. The wavefront spreading is consistent with $r \sim t^{1/z}$. However finite size effects also cause $P_{ij}(t)$ to saturate to an infinite time form which decays as a power law $\lim_{t \to \infty }\bar{P}(r,t) \sim r^{-\beta}$, similar to the QP-Ising and Zero-wandering cases. The asymptotically spreading Aubry-Andr\'{e} cases can be distinguished from the localised QP-Ising and Zero-wandering cases by verifying that the finite-$q$, infinite time form of $\lim_{t \to \infty } P_{ij}(t)$ is increasingly delocalised as the finite size length scale $q$ is increased, and hence there is unbounded wave-packet spreading in the $q \to \infty$ limit. 

We verify that the power law decay of $\lim_{t \to \infty } P_{ij}(t)$ is genuine for the Zero-wandering and QP-Ising cases, but is a finite size effect in the Aubry-Andr\'{e} case by considering the behaviour of the von-Neumann entropy $S_1$ and 2nd Renyi entropy $S_2$ as $q$ is increased where
 \begin{align}
     S_1 &= - \lim_{t \to \infty} \cexp{ \sum_r P_{i,i+r}\log P_{i,i+r} }_i \\
     S_2 &= - \lim_{t \to \infty} \cexp{ \log \sum_r P_{i,i+r}^2 }_i.
     \label{eq:entropy}
 \end{align}
The behaviours of $S_1$ and $S_2$ with increasing $q$ allow us to distinguish three cases
\begin{itemize}
    \item \emph{Delocalised spectrum}: 
    If the \emph{entire} spectrum is delocalised, $\lim_{t \to \infty} P_{ij}(t)$ is spread over increasingly many sites and both $S_1$ and $S_2$  grow asymptotically without bound. This is seen for both the Aubry-Andr\'{e} case (Fig~\ref{fig:Entropy}, lower panel, blue data) and for the Ising case when there is no mobility edge (Fig~\ref{fig:Entropy}, lower panel, gold data).
    \item \emph{Localised spectrum}: If there are no more than a vanishing fraction of delocalised states $\lim_{t \to \infty} P_{ij}(t)$ saturates to a limiting form, and hence both $S_1$ and $S_2$ saturate to a finite value, this is seen both for the Zero wandering case (Fig~\ref{fig:Entropy}, upper panel, blue data) and for the QP-Ising case (data not shown).
    \item \emph{Finite mobility edge} For a spectrum with a finite localised fraction and finite delocalised fraction, $\lim_{t \to \infty} P_{ij}(t)$ has a component that saturates, and a component that spreads, hence $S_2$ saturates whereas $S_1$ grows without bound. This is seen for the Ising case when there is a mobility edge (Fig~\ref{fig:Entropy}, upper panel, gold data).
\end{itemize}

\section{Discussion}

Weak quasi-periodic modulation is perturbatively irrelevant at the clean Ising transition~\cite{Luck:1993ad}. However, sufficiently strong QP modulation, or QP sequences destabilize this transition and drive the TFIM to a new QP Ising fixed point. The critical properties of this fixed point are found to be intermediate to the clean and randomly disordered cases. 
We have focussed on two specific conjectures of Ref~\cite{crowley2018quasi}, detailed in Sec.~\ref{sec:intro}, and have presented evidence that they generically hold. We have additionally shown that with fine tuning either of these conjectures may be violated.

The second conjecture posited the equality of the dynamical exponent and the localisation exponent $z=z_\mathrm{L}$. In randomly modulated, and generic QP modulated transitions the localisation of excitations and change to universality class are concomitant~\cite{fisher1992random,fisher1995critical,fisher1999phase,crowley2018quasi}, and supports the idea that they are necessarily related. However, as we show modulation can induce modified critical scaling without localising excitations (as for $\Delta = 0$), while Ref.~\cite{crowley2018quantum} shows correlated modulation may localise excitations without altering critical scaling, it follows that these two phenomena may be fully decoupled (see Tab.~\ref{tab:ModIsing}). 
This has consequences for the dynamics of correlation functions, as the delocalisation of excitations in the Aubry-Andr\'e case ($\Delta = 0$) allows the operator spreading to continue without bound.

It is straightforward to realize QP modulation in optical experiments by introducing multiple lasers with incommensurate wavelengths~\cite{roati2008anderson,deissler2010delocalization,schreiber2015observation,bordia2016coupling,luschen2017signatures,dal2003light,fallani2007ultracold,lahini2009observation,modugno2010anderson,segev2013anderson}. 
The harder experimental element in such contexts is the preservation of an effective Ising symmetry. Possible host systems include: chains of trapped ions with hyperfine degrees of freedom~\cite{smith2016many,qiong2015parity}; Rydberg ions trapped in optical tweezers~\cite{glaetzle2015designing,labuhn2016tunable}; the staggering transition of ultracold atoms~\cite{simon2011quantum}; or the zig-zag transition in trapped ions~\cite{enzer2000observation,shimshoni2011quantum}.

Though the aforementioned technological developments in optical experiments have driven a recent interest in smooth QP modulation~\cite{iyer2013many,ganeshan2015nearest,varma2017fractality,Chandran:2017ab,gopalakrishnan2017self,setiawan2017transport,crowley2018quasi,szabo2018non}, there is a more longstanding interest in QP models~\cite{satija1988quasiperiodic,you1991quantum,vidal1999correlated,hermisson2000aperiodic,hida2001quasiperiodic,tong2002quantum,hida2005renormalization,vieira2005low,vieira2005aperiodic} originally motivated the discovery and growth of quasicrystals~\cite{shechtman1984metallic,levine1984quasicrystals,merlin1985quasiperiodic}. These systems do not naturally realise smooth QP modulation, but rather QP sequences, as described in Sec.~\ref{sec:zw}. These are captured in our analysis by choosing $\J_j = \J(Qj)$, $\h_j = \h(Q j)$ with $\J(\theta), \h(\theta)$ as piece-wise constant $2\pi$ periodic functions. Ising chains modulated by QP sequences have logarithmic wandering, and hence (by the conjectures of Ref.~\cite{crowley2018quasi} confirmed here) critical scaling described by the QP-Ising universality. However, as these Ising chains have no small couplings we expect them to have fully delocalised spectra formed of multi-fractal excitations. This is consistent with our observations, and the findings of previous studies in free particle models modulated by QP sequences~\cite{hiramoto1988dynamics,hiramoto1992electronic,ketzmerick1997determines}. Thus, we refine the conjecture of Ref.~\cite{crowley2018quasi} in the case of modualtion with QP sequences. At the Ising transition we conjecture these models to have the same critical properties and phenomenology as the Aubry-Andr\'e case studied in this manuscript, that is, critical exponents set by the wandering coefficient $w$ only, with delocalised excitations and hence no localisation length exponent $z_\mathrm{L}$. We lastly note that the discussion here includes the full class of QP sequences, and furthermore any modulation generated by discontinuous $\J(\theta),\h(\theta)$, which are all captured using our methodology. This extends previous work~\cite{doria1988quasiperiodicity,igloi1988quantum,tracy1988universality,ceccatto1989quasiperiodic,kolavr1989attractors,benza1989quantum,benza1990phase,lin1992phase,Luck:1993ad,turban1994surface,grimm1996aperiodic,hermisson1997aperiodic,igloi1997exact,igloi1998random,hermisson1998surface,oliveira2012strong,yessen2014properties} which has been restricted to QP sequences satisfying an inflation rule.

\begin{acknowledgments}
We are grateful to D. Speyer for useful correspondence on the calculation of Eq.~\eqref{eq:WQ} (see Ref.~\cite{speyerprivate}). 
We thank B. Altshuler, Y.Z. Chou, M. Foster, S. Gopalakrishnan, D. Huse, B. McCoy, J.H. Pixley, M. Shumovskyi, S. Sondhi and V. Varma for useful discussions, and the Shared Computing Cluster (administered by Boston University Research Computing Services) for computational support. 
A.C. and C.R.L. acknowledge support from the Sloan Foundation through Sloan Research Fellowships and from the NSF through grants DMR-1752759 (A.C.) and PHY-1752727 (C.R.L.).
\end{acknowledgments}

\bibliography{long-quasi-bib}

\appendix

\section{Relation of group velocity to bandwidth}
\label{app:group_vel}

In this section we show the result that 
\begin{equation}
    W = \pi \bar{v}
    \label{eq:wpiv}
\end{equation}
where $\bar{v}$ is the mean absolute group velocity, and $W$ the total bandwidth, these are given respectively by
\begin{align}
    \bar{v} & = \frac{1}{q} \sum_\alpha  \int_{-\pi/q}^{\pi/q} \frac{d k}{2 \pi/q} |\partial_k \epsilon_\alpha(k)| \\
    W &= \sum_\alpha \left( \max_k \epsilon_\alpha(k) - \min_k \epsilon_\alpha(k) \right).
\end{align}
We show below (Sec.~\ref{sec:BandExtrema}) that each band has exactly one maximum and one minumum, from this it follows that 
\begin{equation}
    \int_{-\pi/q}^{\pi/q} d k |\partial_k \epsilon_\alpha(k)| = 2 \left( \max_k \epsilon_\alpha(k) - \min_k \epsilon_\alpha(k) \right)
\end{equation}
and \eqref{eq:wpiv} follows.

\subsection{Band extrema}
\label{sec:BandExtrema}

The excitation mode energies $\epsilon_\alpha(k)$ are the roots of the characteristic polynomial $\chi(\epsilon_\alpha,k)=0$, where 
\begin{align}
    \chi(\epsilon,k) &= |\mathcal{H}(k) - \epsilon| = \prod_{\alpha = 1}^q \left( \epsilon_\alpha^2(k) - \epsilon^2 \right) = \sum_{n=1}^q \chi_{2n} \epsilon^{2n}.
\end{align}
All coefficients $\chi_{2n}$ are independent of $k$ for $n>0$. Thus all the dependency on $k$ comes from $\chi_0$
\begin{equation}
\begin{aligned}
    \chi_0 &= (-1)^q \left| \prod_i \h_i - \mathrm{e}^{- i k q} \prod_i \J_i \right|^2 \\
    & = (-1)^q \left[ P_\h^2 + P_\J^2 - 2 P_\h P_\J \cos(k q) \right]
\end{aligned}
\end{equation}
for $P_\h = \prod_i \h_i$, $P_\J = \prod_i \J_i$. Thus $\chi_0$ has extrema at $k q = 0 ,\pi$ and changes monotonically between them. As 
\begin{equation}
    \pdev{\epsilon_\alpha (k)}{k} = \left. \left. \pdev{\chi}{\epsilon} \right|_{\epsilon = \epsilon_\alpha (k)} \middle/ \pdev{\chi_0}{k} \right.
\end{equation}
we see that $\partial_k \epsilon_\alpha (k)$ changes sign only where $\partial_k{\chi_0}$ changes sign, and hence each band has exactly two extrema. Here we have used that $\partial_\epsilon{\chi}$ does not change sign as $k$ is varied
\begin{equation}
    \mathrm{sign} \left( \left. \pdev{\chi}{\epsilon} \right|_{\epsilon = \epsilon_\alpha (k)} \right) = (-1)^{q-\alpha}
\end{equation}
where $\alpha = 1 \ldots q$ indexes the positive roots from smallest to largest.

\section{Spectral measure of tri-diagonal matrices}
\label{app:spec_bound}

We prove the bound
\begin{equation}
W = \sum_\alpha W_\alpha \leq 2 \pi \min_i (|J_i|,|h_i|) \sim q^{-1}.
\label{eq:spectralbound}
\end{equation}
where $W_\alpha = \max_k \epsilon_\alpha(k) - \min_k \epsilon_\alpha(k)$ is the width of the $\alpha$th band.  is the total width of the $\alpha$th band of $\mathcal{H}$. This bound is trivially generalisable to any tri-diagonal matrix.

The momentum appears as a phase~$\e^{i k q}$ gained on hopping a distance $q$. Without loss of generality we choose a gauge in which the phase appears entirely on $\J_{\min} = \min(|\J_i|,|\h_i|)$, the smallest magnitude coupling of either form.

As we showed in Sec.~\ref{sec:BandExtrema} that the $k$-dependence of the characteristic polynomial $|\mathcal{H}-\epsilon|=0$ is entirely in a simple cosine dependence of constant term $\chi_0=|\mathcal{H}|$. A consequence of this is that each band $\epsilon_\alpha(k)$ has two stationary points, which lie at $k = 0,\pi/q$, with $\epsilon_\alpha(k)$ changing monotonically between them. Thus it follows 
\begin{equation}
\begin{aligned}
\sum_\alpha W_\alpha &= \sum_\alpha \left|\epsilon_\alpha(\pi/q)-\epsilon_\alpha(0)\right| \\
& = \sum_\alpha \left|\int_0^{\pi/q} \d k \, \partial_k \epsilon_\alpha \right| \\
& = \sum_\alpha \int_0^{\pi/q} \d k \, \left| \partial_k \epsilon_\alpha\right| .
\end{aligned}
\end{equation}
From first order perturbation theory $\partial_k \epsilon_\alpha = \qexp{\partial_k \mathcal{H}}{\epsilon_\alpha}$. Which yields
\begin{equation}
\begin{aligned}
\sum_\alpha W_\alpha &= \sum_\alpha \int_0^{\pi/q} \d k \left|\qexp{\partial_k \mathcal{H}}{\epsilon_\alpha} \right| \\
& \leq  \int_0^{\pi/q} \d k |\partial_k \mathcal{H}|_1 \\
\end{aligned}
\label{eq:specbound2}
\end{equation}
where $|A|_1 = \tr{\sqrt{AA^\dag}}$ denotes the Ky Fan norm. The equality in~\eqref{eq:specbound2} follows from the fact that $\ket{\epsilon_\alpha}$ forms a complete basis. This can be seen explicitly by using the eigen-decomposition $\partial_k \mathcal{H} = \sum_\lambda \ket{\lambda}\lambda\bra{\lambda}$. 
\begin{equation}
\begin{aligned}
\sum_\alpha \left|\qexp{\partial_k \mathcal{H}}{\epsilon_\alpha} \right| &= \sum_{\alpha} \left| \sum_\lambda \left|\braket{\lambda}{\epsilon_\alpha}\right|^2 \lambda \right|.
\\
& \leq \sum_{\alpha,\lambda} |\braket{\lambda}{\epsilon_\alpha}|^2 |\lambda| \\ & = \sum_\lambda |\lambda| \\ &= |\partial_k \mathcal{H}|_1.
\end{aligned}
\end{equation}

The final step is to show 
\begin{equation}
|\partial_k \mathcal{H}|_1 = 2 q \min_i(|J_i|,|h_i|).
\label{eq:KFdH}
\end{equation}
This follows from our gauge choice, in which we put the phase exclusively on the smallest coupling $\J_{\min}=\min_i(|J_i|,|h_i|)$. Thus $\partial_k \mathcal{H}$ is a $q \times q$ matrix with two non-zero elements, one, $i q \J_{\min} \e^{i k q}$, on the first-diagonal, and its conjugate $-i q \J_{\min} \e^{-i k q}$ on the first-sub-diagonal. This matrix has two eigenvalues $\pm q \J_{\min}$, and so its Ky Fan norm is $|\partial_k \mathcal{H}|_1 =2  q \J_{\min}$ for all $k$, so~\eqref{eq:KFdH} and hence~\eqref{eq:spectralbound} follows via~\eqref{eq:specbound2}.

\section{Numerically extracted $z$ for smooth modulation with $Q/2\pi$ a metallic mean}

\begin{figure}
\begin{center}
\includegraphics[width=0.47\textwidth]{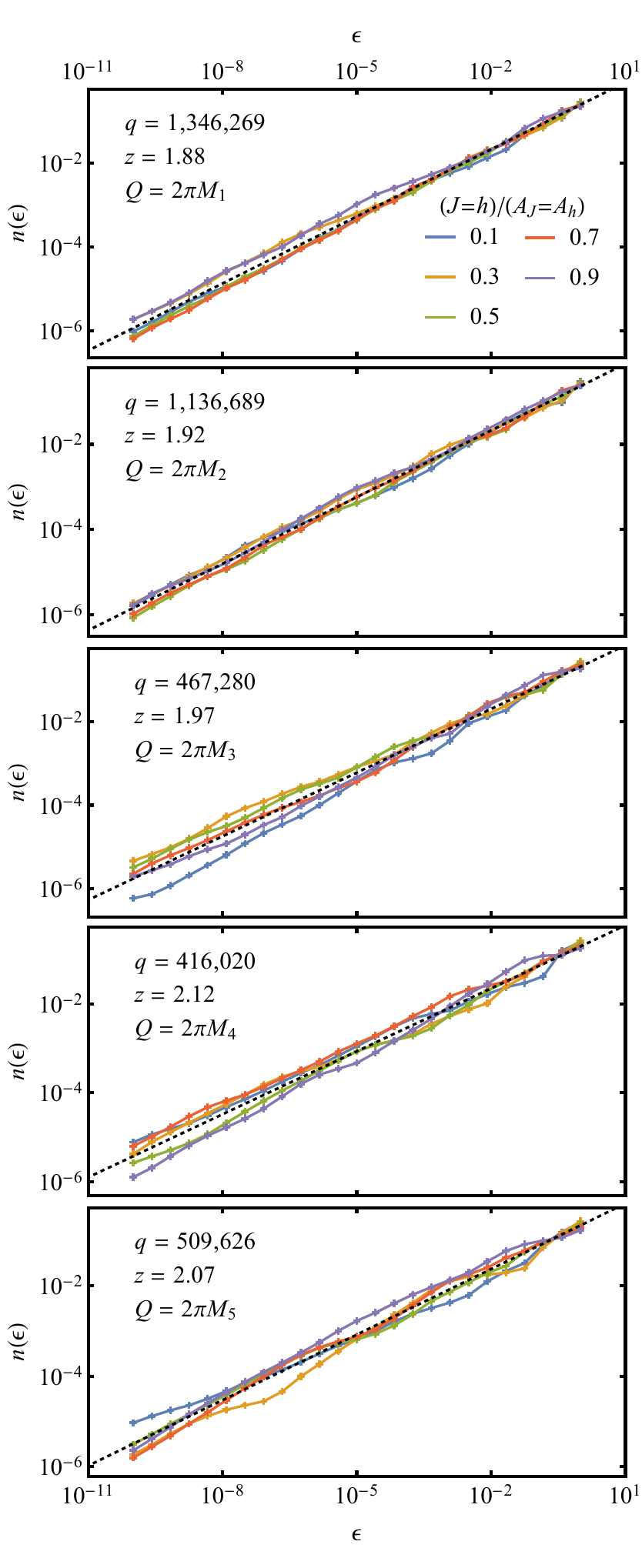}
\caption{
\emph{DOS for Smooth modulation with different $Q$:} Data is shown (solid colours) for the integrated DOS $n(\epsilon)$ for $Q/2\pi = M_n$ for $n=1 \ldots 5$ (see Eq.~\eqref{eq:metallics}). Couplings~\eqref{eq:SineMod} used with values of $(\J=\h)/(\AJ=\Ah)$ shown in legend, value of $q$ and $Q$ inset. Fit line $n(\epsilon) \sim \epsilon^{1/z}$ (dotted black) with value of $z$ inset. Data is averaged over $\phi,\Delta$, statistical error on the mean is smaller than point size.
}
\label{Fig:smoothDOS_apps}
\end{center}
\end{figure}

\label{App:sineDOS}

Fig.~\ref{Fig:smoothDOS_apps} shows additional data for the integrated density of states for sinusoidal modulation~\eqref{eq:SineMod} from the lower critical line $BC$, for different values of $Q$. We study $Q/2\pi = M_n$, where 
\begin{equation}
\begin{aligned}
    M_n & \equiv \frac{n+\sqrt{n^2+4}}{2} \\
    & = n + \frac{1}{n + \frac{1}{n + \frac{1}{n + \ldots}}},
\end{aligned}
\label{eq:metallics}
\end{equation}
are the `metallic means', and $M_1 \equiv \tau$ is the golden ratio. The values of $z$ extracted from this data are shown in Fig.~\ref{Fig:wzPlot}. As in the main text, we calculate $n(\epsilon)$ using the method of Refs.~\cite{schmidt1957disordered,eggarter1978singular}.

\section{Wandering analysis for square waves}
\label{app:SqWave}

In this appendix we calculate the logarithmic wandering coefficient $w$ for square wave modulation, comment on some previous results, and compare calculations with the estimate $z \approx 1 + w$.

We consider the square wave modulation $\J_j = \J(Qj)$, $\h_j = \h(Qj)$
\begin{align}
\J(\theta) &= \J + \AJ \Pi_{D}(\theta + Q/2 + \phi) \nonumber \\
\h(\theta) &= \h + \Ah \Pi_{D}(\theta + \phi + \Delta)
\label{eq:SQwavemod}
\end{align}
where $\Pi_D(\theta)$ is a $2\pi$ periodic square wave with duty cycle $0<D <1$
\begin{equation}
    \Pi_D(\theta) = \begin{cases}
    1 & 0< \theta \leq 2 \pi D \\
    0 & 2 \pi D < \theta \leq 2 \pi  ,
    \end{cases}
\end{equation}
This yields couplings which are drawn from the two value \emph{alphabets} $\J_j \in \{\J,\J+\AJ\}$ and $\h_j \in \{\h,\h+\Ah\}$. The results of this analysis similarly will generalise to general~\emph{discontinuous} $\J(\theta)$, $\h(\theta)$. We note that the previously studied cases of generalised Fibonacci sequences~\cite{tracy1988universality,benza1990phase,grimm1996aperiodic,hermisson1997aperiodic,igloi1997exact,igloi1998random,hermisson1998surface} are special cases of~\eqref{eq:SQwavemod}. We take $\J,\h,\AJ,\Ah>0$ without loss of generality.

\subsection{Square wave wandering coefficient $w$}
\label{app:wSq}

\begin{figure}
\begin{center}
\includegraphics[width=0.47\textwidth]{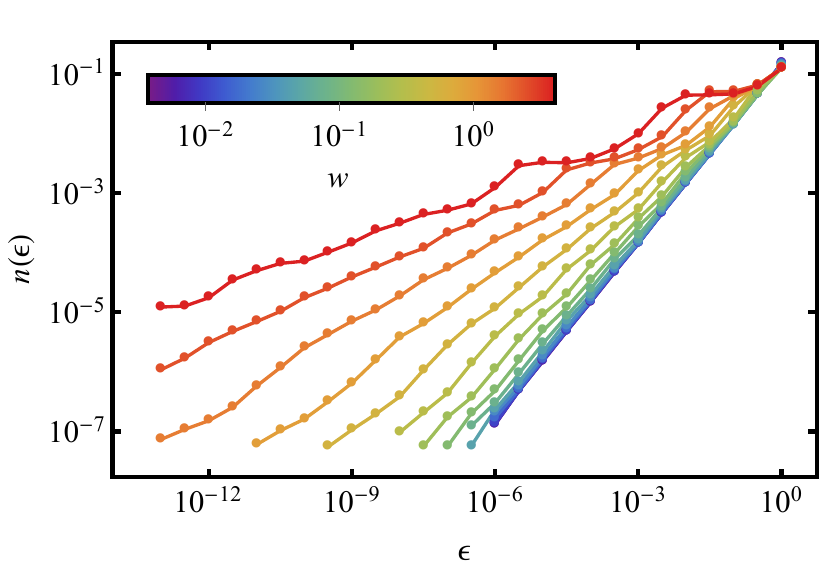}
\includegraphics[width=0.47\textwidth]{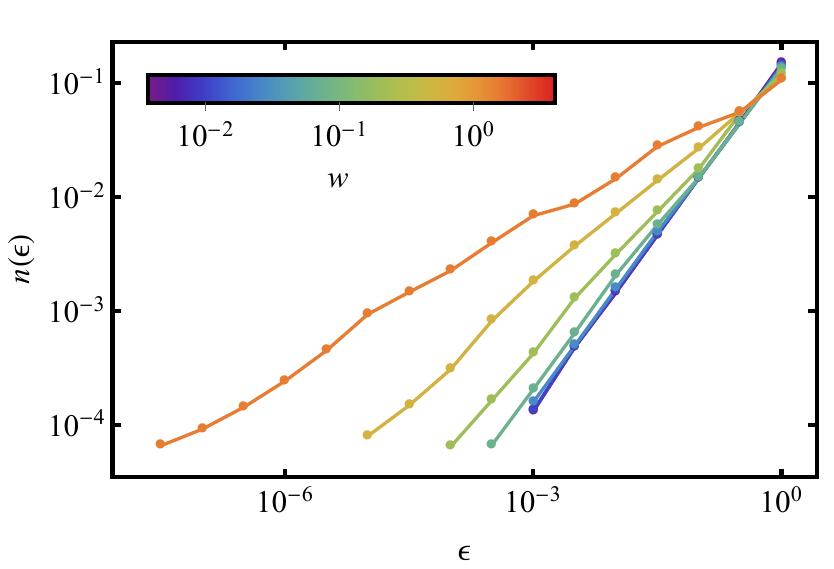}
\caption{
\emph{Square wave density of states:} The integrated DOS $n(\epsilon)$ is plotted for square wave modulation~\ref{eq:SQwavemod} various values of $(\J=\h)/(\AJ=\Ah)$. $n(\epsilon)$ is averaged over $\phi$ and $\Delta$. Each line is coloured according to the value of $w$ (legend inset). $Q/2\pi = M_1 =  (1+\sqrt{5})/2$ (upper plot), $Q/2\pi = M_2 =  1+\sqrt{2}$ (lower plot). 
Error bars not shown; as with the sinusoidal case, fluctuations are deterministic and do not average out. Parameters: $D = 1/\mathrm{e}$, and $q = 1,346,269$ (upper plot), $q=1,136,689$ (lower plot).
}
\label{Fig:DOSSqWave}
\end{center}
\end{figure}

For continuous $J(\theta),h(\theta)$, $w$ is independent of the energetic scales of the model. 
In contrast for $J(\theta), h(\theta)$ with jump discontinuities, one finds $w$ depends explicitly on the modulation amplitude. 
Repeating the calculation of $w_\delta$ for square wave modulation one finds
\begin{align}
    \hat\delta_k &= \frac{\e^{i k \pi D}\sin(\pi k D)}{\pi k} \left( \e^{-i k \Q/2} \log\left|1+\frac{\AJ}{\J}\right| \right. \nonumber \\
    & \quad \quad \quad \left. -   \e^{- i k \Delta} \log\left|1+\frac{\Ah}{\h}\right| \right)
\end{align}
which yields 
\begin{align}
    \cexp{f}_{\text{Ces\`{a}ro}} & = \lim_{k \to \infty} \frac{1}{k}\sum_{k'=1}^k k'^2 |\delta_{k'}|^2 \\
    & = \frac{1}{2\pi^2} \left( \log^2\left|1+\frac{\Ah}{\h}\right| + \log^2\left|1+\frac{\AJ}{\J}\right|\right).
\end{align}
This quantity appears in Eq.~\eqref{eq:w_factor}, and otherwise the calculations proceed as in the the main text. The key difference being that $\cexp{f}_{\text{Ces\`{a}ro}}$ and $w = \cexp{f}_{\text{Ces\`{a}ro}} w_\Q$ now have parameteric dependence on the energy scales $J,h,\AJ,\Ah$.

\subsubsection{Special case: $\Delta = Q(\mathbb{N}+1/2)$}

In this case the wandering is zero $w = 0$. This follows from the same arguments as the sinusoidal case in the main text, and was previously noted for $\Delta = Q/2$ in Ref.~\cite{igloi1988quantum}.

In the sinusoidal case, which is similarly Harris-Luck marginal, the presence of small couplings nonetheless leads to an altered dynamical exponent. Here in the corresponding square wave case, there are no small couplings. 
That is, $\min J_j$, $\min h_j$ do not scale with the finite size length scale $q$ and the modulation is Ising irrelevant.

\subsubsection{Special case: $2\pi D =Q(\mathbb{N}+1/2)$}
\label{app:specialD}

We note there is a corresponding dependence on special values of $D$, analogous to the special values of $\Delta$. E.g. we notice if $D = Q/2\pi$ that 
\begin{align}
    \hat\delta_k &= \frac{\sin(k Q/2)}{\pi k} \left( \log\left|1+\frac{\AJ}{\J}\right| \right. \nonumber \\ 
    & \quad \quad \quad \quad \left. - \e^{i k (Q/2- \Delta)} \log\left|1+\frac{\Ah}{\h}\right| \right)
\end{align}
leading to an exact cancellation with the denominator of~\eqref{eq:wsum} and hence $w=0$. Such an exact cancellation occurs for all $D = n Q/2\pi + m \pi$ for $n,m \in \mathbb{N}$. A previously studied instance of this exact cancellation is if the $h_i$ and $J_i$ follow the Fibonacci word, which is known to be Ising irrelevant~\cite{doria1988quasiperiodicity,igloi1988quantum,ceccatto1989quasiperiodic,kolavr1989attractors,benza1989quantum,benza1990phase,Luck:1993ad,grimm1996aperiodic,hermisson1997aperiodic,igloi1997exact,igloi1998random,hermisson1998surface}.

\subsubsection{Numerically extracted $z$ for square waves}

Fig.~\eqref{Fig:DOSSqWave} shows numerically values of $z$ for Square wave modulation with $Q/2\pi = M_1, M_2$ (see.~\eqref{eq:metallics}). As in the main text, we calculate $n(\epsilon)$ using the method of Refs.~\cite{schmidt1957disordered,eggarter1978singular}.

\end{document}